\documentclass{article}

\usepackage{arxiv}

\usepackage[utf8]{inputenc} 
\usepackage[T1]{fontenc}    
\usepackage{hyperref}       
\usepackage{url}            
\usepackage{pdflscape}
\usepackage{longtable}
\usepackage{booktabs}       
\usepackage{array}
\usepackage{amsfonts}       
\usepackage{nicefrac}       
\usepackage{microtype}      
\usepackage{amsmath}
\usepackage{cleveref}       
\usepackage{lipsum}         
\usepackage{graphicx}
\usepackage[square,sort,comma,numbers]{natbib}
\usepackage{doi}
\usepackage{float}
\usepackage{tabularx}
\usepackage{subcaption}
\usepackage{caption} 
\usepackage{algorithm}
\usepackage{algpseudocode}
\usepackage{rotating} 
\usepackage{ragged2e}

\newcolumntype{C}[1]{>{\centering\arraybackslash}p{#1}} 

\newcolumntype{Y}{>{\RaggedRight\arraybackslash}X}
\newcolumntype{C}[1]{>{\centering\arraybackslash}p{#1}}
\newcolumntype{P}[1]{>{\RaggedRight\arraybackslash}p{#1}}

\newcommand{\clip}{\operatorname{clip}}
\newcommand{\indicator}{\mathbb{I}}

\title{A Reproducible UAV-Assisted VANET Dataset Generator for Fragmentation Risk Analysis in Intelligent Transportation Systems}

\newif\ifuniqueAffiliation
\uniqueAffiliationtrue

\ifuniqueAffiliation 
\author{ \href{https://orcid.org/0009-0002-7235-8653}{\includegraphics[scale=0.06]{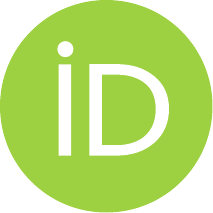}\hspace{1mm}Bappa~Muktar}\thanks{Corresponding author.} \\
	Department of Computer Science\\
	University of Quebec in Outaouais (UQO)\\
	Gatineau, QC J8X 3X7 \\
	\texttt{mukb06@uqo.ca} \\
	\And
	\href{https://orcid.org/0000-0002-9228-4543}{\includegraphics[scale=0.06]{orcid.pdf}\hspace{1mm}Justin~Moskolaï~Ngossaha} \\
	Department of Computer Science\\
	University of Douala \\
	Cameroon, Douala BP 2701 \\
	\And
	\href{https://orcid.org/0000-0003-0324-1443}{\includegraphics[scale=0.06]{orcid.pdf}\hspace{1mm}Adama~Nouboukpo} \\
	Department of Computer Science\\
	University of Quebec in Outaouais (UQO)\\
	Gatineau, QC J8X 3X7 \\
}
\else
\usepackage{authblk}

\newbox{\orcid}\sbox{\orcid}{\includegraphics[scale=0.06]{orcid.pdf}} 
\author[1]{%
	\href{https://orcid.org/0009-0002-7235-8653}{\usebox{\orcid}\hspace{1mm}Bappa~Muktar\thanks{\texttt{mukb06@uqo.ca}}}%
}
\author[1,2]{%
	\href{https://orcid.org/0009-0009-0944-9721}{\usebox{\orcid}\hspace{1mm}Vincent.~Fono\thanks{\texttt{stariate@ee.mount-sheikh.edu}}}%
}
\affil[1]{Department of Computer Science, Cranberry-Lemon University, Pittsburgh, PA 15213}
\affil[2]{Department of Electrical Engineering, Mount-Sheikh University, Santa Narimana, Levand}
\fi

\renewcommand{\shorttitle}{UAV-Assisted VANET Dataset Generator}

\hypersetup{
pdftitle={A template for the arxiv style},
pdfsubject={q-bio.NC, q-bio.QM},
pdfauthor={David S.~Hippocampus, Elias D.~Striatum},
pdfkeywords={First keyword, Second keyword, More},
}

\begin{document}
\maketitle


\begin{abstract}
	Vehicular Ad Hoc Networks (VANETs) are a key component of Intelligent Transportation Systems, enabling cooperative communication among vehicles and between vehicles and roadside infrastructure. However, their highly dynamic topology makes them vulnerable to network fragmentation, particularly in highway scenarios, low-density traffic conditions, localized accident zones, and communication-stressed environments. Although Unmanned Aerial Vehicles (UAVs) have been increasingly investigated as temporary aerial relays for improving VANET connectivity, reusable, future-labeled, and reproducible datasets designed to support short-term fragmentation risk analysis remain limited. This paper proposes a reproducible UAV-assisted VANET dataset generator for short-term fragmentation risk prediction. The proposed framework simulates a two-lane highway scenario in which vehicles move in opposite directions while UAVs operate as aerial support nodes. It incorporates multiple data collection profiles, including free-flow traffic, localized accidents, sparse extended topologies, dense bursty traffic, and mixed stress conditions. During each simulation episode, the generator periodically extracts mobility, topology, UAV coverage, and communication-window features, then assigns each sample a future fragmentation label based on the network state observed after a configurable prediction horizon. An illustrative generated dataset is descriptively characterized in terms of scenario balance, UAV policy balance, future-label distribution, scenario-specific label behavior, and representative feature ranges. By providing a modular, extensible, and reproducible ns-3-based data-generation framework, this work offers a practical basis for future supervised learning studies and connectivity management strategies in UAV-assisted VANETs.
\end{abstract}


\section{Introduction} \label{sect:s1}
Intelligent Transportation Systems (ITS) increasingly rely on connected, cooperative, and data-driven mobility services to improve road safety, enhance traffic efficiency, and strengthen emergency response capabilities. In this context, vehicles, roadside infrastructure, and communication networks are expected to interact in a coordinated, efficient, and dynamic manner. Vehicular Ad Hoc Networks (VANETs) play a central role in this ecosystem by enabling information exchange among vehicles through Vehicle-to-Vehicle (V2V) communications, as well as between vehicles and roadside infrastructure through Vehicle-to-Infrastructure (V2I) communications, without necessarily depending on fixed cellular infrastructure. Recent studies on 5G V2X (Vehicle-to-Everything) architectures emphasize that connected vehicles, emergency vehicles, and cooperative driving services require reliable, robust, and low-latency communications, particularly in road environments characterized by highly dynamic mobility patterns \cite{pawar2024}. However, maintaining stable communication links in VANETs remains a major challenge under real-world conditions, as the network topology continuously evolves under the influence of factors such as vehicle density, relative speed, radio transmission range, lane distribution, and the occurrence of road events \cite{shahwani2022}.

Network fragmentation is one of the most critical challenges in VANETs. Unlike conventional wireless networks, whose topology is generally more stable, VANETs exhibit connectivity patterns that strongly depend on vehicle mobility, spatial distribution, and traffic conditions. As a result, network connectivity may deteriorate rapidly when vehicles move away from each other, when communication links are disrupted by mobility dynamics, or when road incidents cause localized slowdowns and uneven vehicle distribution. From a communication perspective, network fragmentation directly compromises VANET performance by reducing packet delivery ratio, increasing end-to-end delay, degrading routing stability, and limiting the availability of safety-critical messages. Connectivity analysis therefore constitutes a fundamental aspect of the design, optimization, and evaluation of VANETs, particularly when these networks are intended to support safety-critical applications. Previous studies have shown that preserving connectivity in VANETs does not depend solely on topological proximity among nodes, often modeled using graph-based representations, but also on physical-layer constraints, radio link quality, and Quality of Service (QoS) requirements \cite{neelakantan2013}. This issue becomes particularly pronounced on highways and in low-density traffic areas, where communication infrastructure may be limited and inter-group distances between vehicles may exceed the effective radio transmission range, thereby compromising network connectivity.

Unmanned Aerial Vehicles (UAVs) have recently emerged as a promising support technology for VANETs. Owing to their aerial mobility, deployment flexibility, and ability to operate as temporary relay nodes, UAVs can complement fixed roadside units, interconnect isolated groups of vehicles, and enhance communication coverage in areas where terrestrial connectivity is insufficient. In this context, UAV-assisted VANET routing has been investigated as a means of providing a broader view of connected road segments and improving packet forwarding decisions \cite{ashraf2024}. Recent studies further confirm that the integration of UAVs into VANETs represents a rapidly growing research direction in the field of ITS. This convergence enables a wide range of applications, including routing support, emergency communications, flexible extension of communication infrastructure, and the provision of advanced mobility services in smart city environments \cite{ashraf2024}. In addition, several UAV-assisted deployment strategies have been proposed to improve coverage in road segments characterized by low traffic density or limited connectivity, particularly when fixed roadside units are unable to provide sufficient communication coverage \cite{jain2024}.

Despite these advances, most existing studies on UAV-assisted VANETs remain primarily focused on protocol design, routing performance optimization, resource allocation, and communication infrastructure deployment. By contrast, the construction of reusable, future-labeled, and reproducible datasets aimed at identifying early indicators of VANET fragmentation remains relatively underexplored. This gap is significant because fragmentation is not only an instantaneous connectivity state, but also a predictive problem. A VANET may appear partially connected at a given time, while evolving mobility patterns, traffic bursts, localized slowdowns, or UAV positioning may indicate imminent fragmentation a few seconds later. The ability to anticipate such degradation would enable a UAV-assisted VANET management framework to reposition UAVs proactively, dynamically adjust communication policies, and activate preventive relay mechanisms before packet transmission performance reaches a critical level.

Machine learning provides a relevant approach for addressing this prediction problem. In ITS, machine learning is increasingly used to model complex relationships among mobility, traffic conditions, communication performance, and contextual road-environment variables \cite{yuan2022}. However, supervised learning models require datasets that are sufficiently diverse, properly labeled, and reproducible. For VANET fragmentation prediction, a relevant dataset must include not only raw communication indicators such as throughput, delay, and packet delivery ratio, but also topological and spatio-temporal descriptors, including the number of connected components, the largest connected component ratio, the average node degree, inter-vehicle spacing, UAV coverage ratio, and traffic evolution over successive temporal windows.

Existing simulation-based studies generally report performance results for a specific protocol, configuration, or scenario, without necessarily providing a modular, reproducible, and reusable data generation framework. This limitation reduces comparability across studies and makes it difficult to assess the generalization capability of proposed prediction models beyond manually configured simulations. Simulation nevertheless remains a practical and scientifically relevant approach for generating VANET datasets, especially when large-scale real-world traces of connected vehicles are unavailable, costly to collect, or difficult to instrument. The ns-3 simulator is widely used for packet-level network simulation and provides a flexible framework for modeling wireless communications, routing protocols, network applications, and performance indicators \cite{riley2010}. Similarly, SUMO is commonly used to model road traffic and vehicle mobility at the microscopic level, making it particularly suitable for coupling transportation dynamics with network behavior \cite{behrisch2011}. Together, these tools provide a solid foundation for conducting controlled, reproducible, and configurable VANET experiments. Nevertheless, the scientific value of simulation-based datasets strongly depends on the rigor of scenario design, the relevance of label definition, and the ability of the generated data to support meaningful machine learning tasks rather than merely descriptive performance analysis.

In this context, this paper proposes a reproducible UAV-assisted VANET dataset generator for fragmentation risk analysis in ITS. The proposed generator simulates a two-lane highway scenario in which vehicles travel in opposite directions while UAVs operate as temporary aerial support nodes. The scenario incorporates localized accident zones that may slow down traffic and traffic bursts that may overload the communication layer. The generator periodically samples vehicle mobility, network topology, UAV coverage, and traffic-window characteristics. Each sample is then assigned a future fragmentation label based on the network state observed after a configurable prediction horizon. Accordingly, the generated dataset provides a structured basis for future supervised learning studies aimed at predicting short-term fragmentation risk from the current state of the VANET.

The main contributions of this work are summarized as follows:
\begin{enumerate}
\item  A modular and reproducible ns-3-based UAV-assisted VANET dataset generator is proposed for fragmentation risk analysis. The generator supports the simulation of diverse mobility and communication conditions and produces traces suitable for machine learning tasks. 
\item A feature extraction process is defined to characterize the temporal state of the VANET. This process captures mobility-related, topology-related, UAV-coverage-related, and communication-traffic-related indicators. 
\item  A predictive labeling strategy is introduced to transform simulation traces into a supervised learning dataset. This strategy associates each current network state with a short-term future fragmentation risk. 
\item Several representative scenarios are supported, including free-flow traffic, localized accident conditions, sparse extended topologies, dense bursty traffic, and mixed constraint scenarios combining multiple mobility and communication stress factors. 
\item The proposed framework records both baseline and proactive UAV positioning policies, thereby enabling descriptive comparison and supporting future studies on predictive UAV-assisted fragmentation mitigation in VANETs. 
\item An illustrative generated dataset is descriptively characterized in terms of scenario balance, UAV policy balance, future-label distribution, scenario-specific label behavior, and representative feature ranges, thereby supporting reproducible downstream machine learning experiments. 
\end{enumerate}

The remainder of this paper is organized as follows. Section~\ref{sect:s2} reviews related work on VANET connectivity, UAV-assisted VANETs, and simulation-based datasets for learning-oriented ITS applications. Section~\ref{sect:s3} presents the proposed UAV-assisted VANET dataset generator and its software architecture. Section~\ref{sect:s4} describes the simulation scenario, mobility assumptions, UAV support configuration, communication setup, and data collection profiles. Section~\ref{sect:s5} formalizes the feature extraction and future-labeling methodology. Section~\ref{sect:s6} presents the implementation, reproducibility controls, and dataset generation protocol. Section~\ref{sect:s7} characterizes the dataset generated by the proposed framework, discusses the scientific relevance and limitations of the approach, and concludes the paper.


\section{Related Work} \label{sect:s2}

Recent studies on VANETs have examined connectivity preservation, routing robustness, UAV-assisted communication, simulation-based dataset generation, and learning-oriented prediction or detection tasks. These research directions are directly relevant to fragmentation risk analysis, as the reliability of VANET services depends not only on instantaneous communication performance but also on the ability to anticipate future degradation in network connectivity. Accordingly, this section reviews recent contributions organized into four main categories: VANET connectivity and fragmentation analysis, UAV-assisted VANETs and aerial relaying, dataset generation for ITS, and learning-oriented VANET datasets for risk-related prediction tasks. The section concludes by positioning the proposed generator in relation to the existing literature.
 
  \subsection{VANET Connectivity and Fragmentation Analysis}
 
  This subsection reviews studies that analyze VANET connectivity as a dynamic property influenced by mobility, communication range, infrastructure deployment, and network density. Particular attention is given to works that clarify how link instability, sparse traffic conditions, and topology changes affect the formation of connected components and the continuity of vehicular communication services.
 
  Dutta, A. et al. have conducted a comprehensive review of recent developments in VANETs for traffic monitoring, safety applications, and remote sensing services \cite{dutta2024}. The study addresses the need for reliable vehicular communication in dynamic transportation environments where mobility, routing instability, channel access, and network density affect service continuity. The authors review communication technologies, data acquisition mechanisms, clustering strategies, and routing approaches designed to support Vehicle-to-Everything (V2X) services. Their contribution lies in synthesizing the role of VANETs in safety messaging, traffic monitoring, and limited-coverage scenarios. The study is relevant to the present work because it highlights the importance of connectivity-aware VANET design. However, it remains a survey and does not provide a reusable simulation-based dataset generator for predicting future fragmentation risk.
 
  Pawar, V. et al. have conducted a detailed study on 5G V2X architectures, vehicular use cases, emergency vehicle communication, current challenges, and future research directions \cite{pawar2024}. The work addresses the communication requirements of connected vehicles, cooperative driving services, and safety-critical emergency applications. The authors examine the role of 5G-enabled V2X communication in supporting low latency, reliability, scalability, and high service availability in dynamic road environments. Their contribution is significant because it positions V2X architectures as a foundation for next-generation Intelligent Transportation Systems. This work is relevant to fragmentation analysis because it reinforces the need for stable communication paths in scenarios where emergency vehicles and cooperative mobility services depend on continuous connectivity. Nevertheless, the study does not focus on simulation-based fragmentation prediction or dataset generation.
 
  Okello, K. et al. have conducted a connectivity analysis of VANETs under dynamic communication ranges \cite{okello2025}. The research problem addressed is the limitation of conventional connectivity models that assume fixed radio ranges and ideal communication conditions. The authors propose an analytical framework that accounts for statistical variations in network topology, wireless fading, and obstructions caused by large vehicles. This approach makes connectivity assessment more realistic by integrating the uncertainty of the physical layer into the evaluation of inter-vehicle communication. The study contributes to the understanding of how dynamic communication ranges affect link availability and network connectivity. It is relevant to the proposed work because it shows that VANET fragmentation cannot be analyzed solely through graph proximity; physical-layer constraints and communication reliability must also be considered.
 
  Gu, X. et al. have conducted research on cluster-based Roadside Unit (RSU) deployment for VANETs by integrating communication, sensing, and computing capabilities \cite{gu2024}. The study addresses the challenge of identifying efficient RSU deployment locations in road networks where full infrastructure coverage is costly or impractical. The authors propose a network-based renormalization method that exploits information flow and geographical location to hierarchically determine RSU placement. Their results indicate that hierarchical deployment can improve communication performance compared with selected optimization-based strategies. This work is relevant because it demonstrates that road topology, spatial distribution, and infrastructure placement have direct implications for connectivity preservation. However, the contribution is centered on fixed infrastructure planning, whereas the present work focuses on UAV-assisted dataset generation for short-term fragmentation risk prediction.
  
  Overall, these studies confirm that VANET connectivity is influenced by a combination of mobility, communication range, physical-layer uncertainty, road topology, and infrastructure placement. However, they mainly analyze connectivity as a current network property or as an infrastructure planning issue. They do not provide a reusable simulation framework that converts connectivity evolution into future-labeled samples for short-term fragmentation risk analysis.
 
  \subsection{UAV-Assisted VANETs and Aerial Relaying}
 
  This subsection discusses recent contributions in which UAVs are introduced as flexible aerial relays or complementary communication infrastructure for VANETs. The reviewed works are particularly relevant to the present study because they show how aerial nodes can support connectivity restoration, urgent message dissemination, and service continuity in sparse or disrupted vehicular environments.
 
  Ashraf, M. M. et al. have conducted a comprehensive review of UAV integration with VANETs, focusing on architectures, applications, opportunities, and open communication challenges \cite{ashraf2024}. The study addresses the limitations of ground-based VANET communication, particularly frequent link failures, variable node density, mobility-induced disconnections, and limited transmission range. The authors analyze how UAVs can support VANETs as aerial relays, mobile infrastructure, and flexible communication platforms. Their contribution lies in proposing a taxonomy of UAV--VANET integration approaches and identifying design issues across communication layers. This review is highly relevant to the proposed study because it establishes UAVs as a promising mechanism for improving VANET connectivity. However, it does not develop a reproducible dataset generation framework for forecasting fragmentation risk.

 Zear, A. et al. have conducted a systematic review on network partitioning problems and UAV integration for efficient connectivity restoration \cite{zear2025}. The study addresses the issue of wireless network partitioning, where node mobility, failures, or sparse deployment can divide a network into disconnected components. The authors review partition detection and recovery strategies and analyze how UAVs can reconnect isolated network segments. Their contribution is important because it frames UAVs as dynamic agents capable of restoring connectivity in partitioned networks. This work is directly related to VANET fragmentation, since fragmentation can be interpreted as a partitioning problem in highly mobile vehicular environments. However, the reviewed approaches emphasize recovery after partitioning rather than predictive dataset generation for anticipating future fragmentation before communication performance collapses.

 Bouchrit, L. et al. have conducted research on UAV-assisted urgent alert transmission in VANETs for rural highway environments \cite{bouchrit2023}. The study addresses the difficulty of delivering urgent messages in sparse vehicular networks where direct vehicle-to-vehicle paths may be unavailable. The authors propose a unified UAV--VANET architecture in which UAVs act as relay nodes to support urgent alert dissemination among vehicles. Their methodology emphasizes the role of aerial relays in improving message delivery under limited infrastructure and sparse traffic conditions. The contribution is relevant because it demonstrates that UAVs can mitigate connectivity gaps in highway scenarios. Nevertheless, the study focuses on alert transmission and relay support, while the proposed work targets the generation of labeled datasets for predicting short-term fragmentation risk.

 Sethu Narayanan, K. et al. have conducted research on an integrated VANET--FANET multimodal smart transportation system for delivery applications \cite{sethu2025}. The study addresses the need to combine ground vehicular networks and flying ad hoc networks to support intelligent transportation and logistics services. The authors propose a multimodal communication framework in which vehicular and aerial nodes cooperate to improve service continuity and delivery performance. Their contribution lies in demonstrating the potential of integrated VANET--FANET systems for smart transportation applications. This work is relevant because it reflects the growing convergence between vehicular networks and UAV-supported communication. However, its focus is primarily service-oriented, whereas the proposed study concentrates on a reusable dataset generator for topology-, mobility-, traffic-, and UAV-coverage-based fragmentation risk analysis.

 Ye, M. et al. have conducted research on improving transmission in integrated UAV--Intelligent Connected Vehicle networks using opportunistic approaches \cite{ye2025}. The study addresses communication instability in UAV--assisted vehicular networks caused by high mobility, intermittent links, and selfish node behavior. The authors model the integrated UAV--vehicle network as a vehicular delay-tolerant network and propose an opportunistic transmission strategy that considers node cooperation, encounter probability, centrality, energy consumption, and cache size. Their results show improvements in delivery ratio and average delay compared with benchmark methods. This work is relevant because it treats UAVs as flexible communication support entities under disrupted connectivity. However, it focuses on transmission performance optimization rather than on constructing reusable supervised learning datasets for future fragmentation prediction.
 
 These contributions demonstrate the potential of UAVs to improve connectivity, restore partitioned networks, and support message dissemination in sparse vehicular environments. Nevertheless, most UAV-assisted VANET studies focus on communication performance, routing, or post-disconnection recovery. They rarely address the generation of reusable datasets in which UAV coverage, mobility patterns, and topology evolution are jointly encoded to anticipate future fragmentation.

  \subsection{Dataset Generation for Intelligent Transportation Systems}
 
  This subsection focuses on simulation-based dataset generation for ITS and VANET research. The emphasis is placed on works that combine network simulators, traffic mobility tools, and machine learning pipelines to transform synthetic vehicular traces into reusable datasets for detection, classification, or predictive analysis.

 Rashid, K. et al. have conducted research on an adaptive real-time malicious node detection framework using machine learning in VANETs \cite{rashid2023}. The study addresses the challenge of detecting malicious behavior in highly dynamic vehicular environments where attackers may disrupt communication reliability. The authors use OMNeT++ and SUMO to generate realistic VANET traces and evaluate several machine learning classifiers for malicious node detection. Their contribution lies in combining simulation-based data generation with real-time adaptive detection, demonstrating the value of simulator-derived datasets for VANET security. This work is relevant because it illustrates how mobility and network simulation can support supervised learning tasks. However, its dataset is designed for malicious node detection rather than for predicting topology degradation or future fragmentation.

 Bodkhe, U. et al. have conducted research on an Indian SUMO traffic scenario-based misbehavior detection dataset for connected vehicles \cite{bodkhe2025}. The study addresses the scarcity of public datasets reflecting non-European Intelligent Transportation System contexts. The authors introduce the Ahmedabad SUMO Traffic scenario and construct the AhmST dataset to model false data injection attacks affecting position, heading, and speed in Basic Safety Messages. Their contribution is important because it improves geographical diversity and reproducibility in VANET dataset construction. The study is relevant to the present work because it demonstrates the importance of realistic scenario design and reusable datasets. However, its focus is misbehavior detection, while the proposed generator targets future fragmentation risk using mobility, topology, UAV coverage, and traffic-window descriptors.

 Anjali, T. et al. have conducted research on X-GEVON, an explainable intelligent network for detecting multiple attacks in VANET systems \cite{anjali2025}. The study addresses the need for accurate and interpretable intrusion detection in dynamic VANET environments where multiple attack types may occur. The authors construct datasets using a combination of SUMO, OMNeT++, and Python, and design a computationally aware deep learning architecture based on gated recurrent units with optimized hyperparameters. They also apply explainable AI techniques to interpret model decisions. Their contribution lies in combining dataset generation, deep learning, optimization, and interpretability for VANET attack detection. This work is relevant because it highlights the role of simulator-generated data in learning-based VANET analysis. However, it remains attack-centered and does not address future fragmentation labeling.

 Ercan, S. et al. have conducted research on misbehavior detection for position falsification attacks in VANETs using machine learning \cite{ercan2022}. The study addresses the difficulty of detecting vehicles that intentionally falsify their position information, which can compromise trust, routing, and safety applications. The authors propose a machine learning-based detection mechanism that introduces position-related features to improve the identification of falsified messages. Their contribution lies in showing that carefully engineered mobility and position descriptors can enhance misbehavior detection in VANETs. This work is relevant because it demonstrates the importance of feature design in learning-based vehicular network analysis. However, it focuses on identifying malicious position falsification rather than generating future-oriented labels for connectivity fragmentation prediction.
 
 The reviewed works highlight the importance of simulation tools such as SUMO and OMNeT++ for producing controlled and reusable VANET datasets. However, their generated data are mainly designed for misbehavior, intrusion, or attack detection. In contrast, the proposed generator focuses on topology-aware and UAV-assisted fragmentation risk analysis, where the target label is derived from the future connectivity state rather than from malicious behavior.

 \subsection{Learning-Oriented VANET Datasets and Risk-Related Prediction Tasks}
 
  This subsection reviews learning-oriented VANET datasets and predictive frameworks related to network risk, security, congestion, and communication degradation. Although most existing datasets are designed for attack detection or traffic classification, they provide useful methodological insights into feature engineering, labeling, and evaluation protocols for supervised learning in vehicular networks.
 
  Setia, H. et al. have conducted research on machine learning-driven DDoS attack detection in VANET cloud environments \cite{setia2024}. The study addresses security vulnerabilities that arise when VANETs are integrated with cloud services, particularly the risk of DDoS attacks disrupting connected vehicle applications. The authors propose an architectural framework for capturing and analyzing VANET cloud network flows and use machine learning with fuzzification to detect malicious behavior. Their results report high predictive accuracy, supporting the use of supervised learning for VANET security. The study is relevant because it shows that learning-based models can capture complex traffic patterns in vehicular networks. Nevertheless, it does not consider UAV-assisted connectivity support, graph-based fragmentation indicators, or predictive labels related to future network disconnection.
 
  Polat, O. et al. have conducted research on hybrid AI-powered real-time DDoS detection and traffic monitoring for software-defined VANETs \cite{polat2024}. The study addresses the vulnerability of SD-VANET controllers to DDoS attacks, which may disrupt centralized traffic management and intelligent transportation services. The authors propose SD-VANET\_Guard, a hybrid architecture combining one-dimensional convolutional neural networks and decision trees for real-time attack detection. Their contribution lies in integrating AI-based classification into the SDN control plane to improve the security and responsiveness of VANET monitoring. This work is relevant because it demonstrates the value of real-time learning-based analysis in vehicular networks. However, the research remains centered on cyberattack detection and does not construct datasets for anticipating future fragmentation in UAV-assisted VANET topologies.
 
  Haydari, A. et al. have conducted research on RSU-based online intrusion detection and mitigation for VANETs~\cite{haydari2022}. The study addresses cyberattacks targeting data availability and integrity, including false data injection and stealthy DDoS attacks. The authors propose a centralized RSU-based anomaly detection and mitigation framework capable of identifying and localizing attacks without requiring a fixed attack strategy. Their contribution is significant because it combines detection, localization, and mitigation in an online VANET security framework. The study is relevant because RSUs and monitoring infrastructure are important for maintaining network reliability. However, the work focuses on malicious behavior detection through RSU-based monitoring, while the proposed generator emphasizes UAV-assisted connectivity monitoring and future fragmentation risk labeling.
 
  Gopi, R. et al. have conducted research on intelligent DoS attack detection with congestion control in VANETs \cite{gopi2022}. The study addresses the combined problem of denial-of-service attacks and traffic congestion, both of which can degrade vehicular communication and transportation efficiency. The authors propose an Intelligent DoS Attack Detection with Congestion Control technique that incorporates Teaching--Learning-Based Optimization and a GRU-based detection model. Their contribution lies in linking security detection with congestion-aware control, thereby improving communication reliability under hostile network conditions. This work is relevant to the present study because congestion, traffic bursts, and network load are also important contributors to VANET fragmentation. However, it does not provide a modular generator for creating future-labeled fragmentation risk datasets.

 Elsadig, M. A. et al. have conducted research on a lightweight machine learning model for detecting VANET attacks \cite{elsadig2025}. The study addresses the need for efficient, deployable, and low-complexity security mechanisms suitable for connected vehicle environments. The authors use feature selection and class balancing to improve detection performance while reducing computational overhead. Their contribution lies in designing a lightweight pipeline that can support attack detection in resource-constrained vehicular settings. This work is relevant because it reinforces the importance of dataset preprocessing, feature reduction, and model efficiency in VANET-related prediction tasks. However, the study is oriented toward attack detection rather than topology-aware risk forecasting. The proposed generator extends this direction by producing data for fragmentation prediction rather than only security classification.
 
 Taken together, these studies show that machine learning can support VANET monitoring, anomaly detection, and risk-related decision making. However, the notion of risk is predominantly associated with cyberattacks, malicious behavior, or congestion. To the best of our knowledge, relatively few works explicitly formulate VANET fragmentation as a future-oriented supervised learning problem supported by topology, mobility, UAV coverage, and communication-window features.

  \subsection{Positioning of the Proposed Generator}
  \label{sec2.5}

 The reviewed literature shows that significant progress has been made in VANET connectivity modeling, UAV-assisted communication, simulator-based dataset construction, and machine learning-driven VANET security. Existing works have investigated V2X communication requirements, dynamic connectivity, RSU deployment, UAV-supported relaying, misbehavior detection, DDoS classification, and intrusion detection. However, most contributions remain centered on either protocol optimization, infrastructure deployment, attack detection, or model performance evaluation. Even when simulation-generated datasets are introduced, they are generally designed for cybersecurity tasks such as DDoS detection, Sybil detection, false data injection, or malicious node classification.

 The proposed work differs from these studies in both objective and methodological scope. The primary contribution of this work is a reusable UAV-assisted VANET dataset generator rather than a predictive model itself, with the objective of supporting future fragmentation risk prediction studies. The generator is designed to extract mobility, topology, UAV coverage, and communication traffic-window features from simulation traces and to assign each sample a future fragmentation label according to a configurable prediction horizon. This formulation enables VANET fragmentation to be studied not only as a descriptive connectivity analysis problem, but also as a supervised learning task in which the current UAV-assisted VANET state is used to anticipate short-term fragmentation risk. This orientation supports reproducible experimentation, comparative evaluation of prediction models, and future research on proactive UAV-based fragmentation mitigation. A detailed comparative positioning of the reviewed studies and the proposed generator is provided in Appendix~\ref{app:related_work_positioning}, Table~\ref{tab:appendix_related_work_positioning}.


\section{Proposed UAV-Assisted VANET Dataset Generator} \label{sect:s3}

 This section presents the proposed UAV-assisted VANET dataset generation framework, which is designed to produce reproducible and future-labeled simulation data for short-term fragmentation risk analysis in Intelligent Transportation Systems. The generator is implemented in an object-oriented manner using ns-3 and models a highway VANET in which ground vehicles communicate through ad hoc wireless links while UAVs operate as aerial support nodes. Unlike studies that primarily evaluate a routing protocol or a predictive model after simulation, the proposed framework is designed as a dataset generation pipeline. It repeatedly simulates heterogeneous mobility and communication conditions, extracts mobility, topology, UAV coverage, and traffic-window features, and assigns each observation a future fragmentation label according to a configurable prediction horizon.

 \subsection{System Overview}
 \label{subsec:system_overview}

 The system overview formalizes the proposed generator as a time-varying UAV-assisted vehicular network. This abstraction links the physical mobility of vehicles and UAVs to a graph-based representation of communication connectivity, which serves as the basis for feature extraction, fragmentation analysis, and future risk labeling.

 Let the simulated UAV-assisted VANET at time $t$ be represented as an undirected connectivity graph:
 \begin{equation}
 G(t) = \left(V(t), E(t)\right),
 \label{eq:graph_definition}
 \end{equation}
 where the set of nodes is composed of ground vehicles and UAVs:
 \begin{equation}
 V(t) = V_g(t) \cup V_u(t),
 \label{eq:node_set}
 \end{equation}
 with $V_g(t)$ denoting the set of ground vehicles and $V_u(t)$ denoting the set of UAV support nodes. An edge exists between two nodes \(i\) and \(j\) if their three-dimensional Euclidean distance is less than or equal to the configured effective communication range \(R_{\mathrm{eff}}\):

\begin{equation}
e_{ij}(t) \in E(t) \Longleftrightarrow d^{3D}_{ij}(t) \leq R_{\mathrm{eff}},
\end{equation}

where

\begin{equation}
\begin{aligned}
d^{3D}_{ij}(t)
&=
\Big[
(x_i(t)-x_j(t))^2
+
(y_i(t)-y_j(t))^2 
+
(z_i(t)-z_j(t))^2
\Big]^{1/2}.
\end{aligned}
\label{eq:distance_3d}
\end{equation}

The parameter \(R_{\mathrm{eff}}\) represents the effective communication range used by the generator to construct the connectivity graph. It is configurable and provides a controlled abstraction of the communication neighborhood used for topology-aware feature extraction.

 This graph-based abstraction provides a controllable and reproducible way to derive network connectivity from the simulated mobility state. It also enables the generator to compute fragmentation-related indicators, including the number of connected components, the size of the largest connected component, the average node degree, and the normalized largest connected component ratio.

At each sampling instant \(t_k\), the generator converts the current simulated network state into a structured feature vector:
 \begin{equation}
 \mathbf{x}_{t_k}
 =
 \left[
 \mathbf{x}_{\mathrm{mob}},
 \mathbf{x}_{\mathrm{topo}},
 \mathbf{x}_{\mathrm{uav}},
 \mathbf{x}_{\mathrm{traffic}},
 \mathbf{x}_{\mathrm{context}}
 \right],
 \label{eq:feature_vector}
 \end{equation}
 where $\mathbf{x}_{\mathrm{mob}}$ captures mobility-related descriptors, $\mathbf{x}_{\mathrm{topo}}$ captures graph-level topological metrics, $\mathbf{x}_{\mathrm{uav}}$ describes UAV coverage, $\mathbf{x}_{\mathrm{traffic}}$ summarizes communication behavior over the current sampling window, and $\mathbf{x}_{\mathrm{context}}$ records scenario-level information such as accident activity and traffic stress conditions.

 The supervised learning target is not the current fragmentation state, but the future fragmentation state observed after a configurable prediction horizon $H$:
 \begin{equation}
 y_{t_k} = \mathcal{L}\left(G(t_k + H)\right),
 \label{eq:future_label_general}
 \end{equation}
 where \(\mathcal{L}(\cdot)\) denotes the future-labeling function, which is detailed in Section~\ref{sect:s5}. This formulation enables VANET fragmentation to be represented as a future-oriented prediction problem: given the current state of the UAV-assisted VANET, the generated dataset supports the prediction of short-term future fragmentation risk.

 \subsection{Object-Oriented Architecture}
 \label{subsec:oo_architecture}

 The generator is structured around modular components, each responsible for a specific stage of simulation, feature extraction, or dataset construction. This organization improves maintainability and facilitates extensions such as new mobility models, additional UAV positioning policies, alternative traffic profiles, or modified labeling strategies.
 
 The detailed object-oriented structure of the proposed generator is provided in Appendix~\ref{app:oop_architecture}, Figure~\ref{fig:oop_architecture}. The diagram highlights the central role of \texttt{ScenarioRunner} in coordinating the main simulation, feature extraction, UAV control, and dataset writing components.

 The main orchestration is handled by \texttt{ScenarioRunner}, which configures each simulation episode, creates nodes, installs mobility, configures the wireless network stack, installs traffic applications, schedules periodic sampling, and writes labeled samples. The configuration parameters are centralized in \texttt{SimulationConfig}, while \texttt{RuntimeContext} stores ns-3 nodes, devices, IP interfaces, traffic counters, and extracted snapshots. The feature extraction logic is separated from the simulation control logic, allowing the dataset generator to evolve independently from the network scenario itself.
 
 The remaining components implement the main functional stages of the generator. The \texttt{EpisodeConfigurator} defines the scenario profile and episode-specific parameters, while the \texttt{RandomEngine} ensures reproducible randomization. The \texttt{MobilityManager} installs vehicle and UAV mobility models, and the \texttt{NetworkManager} configures wireless communication, routing, and IP addressing. The \texttt{TrafficManager} installs communication applications and records traffic-window counters. The \texttt{TopologyAnalyzer} constructs the connectivity graph and computes graph-level metrics, whereas the \texttt{FeatureExtractor} aggregates mobility, topology, UAV coverage, traffic, and contextual features into structured snapshots. Finally, the \texttt{UavController} implements UAV positioning policies, and the \texttt{DatasetWriter} exports the generated labeled samples to CSV format.

 \subsection{Overall Dataset Generation Algorithm}
 \label{subsec:generation_algorithm}

 Algorithm~\ref{alg:dataset_generation} summarizes the overall dataset generation process. Each episode corresponds to an independent simulation run with a specific scenario profile and UAV policy. The generator periodically extracts network states during the simulation and then transforms these states into labeled samples by looking ahead over the configured prediction horizon.

\begin{algorithm}[H]
 \footnotesize
 \caption{UAV-assisted VANET dataset generation}
 \label{alg:dataset_generation}
 \begin{algorithmic}[1]
 \Require Number of episodes $N_e$, simulation time $T$, sampling start time $t_0$, sampling period $\Delta t$, prediction horizon $H$, effective range $R_{\mathrm{eff}}$, scenario profile set $\mathcal{P}$, UAV policy set $\Pi$, random seed $s$
 \Ensure CSV dataset $\mathcal{D}$
 \State Initialize the random generator with seed $s$
 \State Open the output CSV file and write the dataset header
 \For{$e = 0$ to $N_e - 1$}
     \State Select scenario profile $p_e \in \mathcal{P}$
     \State Select UAV policy $\pi_e \in \Pi$
     \State Randomize episode parameters according to $p_e$
     \State Create vehicle nodes and UAV nodes
     \State Install two-lane highway mobility
     \State Install UAV mobility models
     \State Configure IEEE 802.11p ad hoc wireless communication
     \State Install AODV routing and assign IPv4 addresses
     \State Install UDP sinks on UAV nodes
     \State Install base and burst UDP traffic sources on vehicle nodes
     \For{each sampling time $t_k$ from $t_0$ to $T$}
         \State Build connectivity graph $G(t_k)$
         \State Extract mobility, topology, UAV coverage, and traffic-window features
         \State Compute the current fragmentation risk score
         \If{the proactive UAV policy is enabled}
             \State Update UAV velocities according to the current risk state
         \EndIf
         \State Store snapshot $\mathbf{x}(t_k)$
     \EndFor
     \For{each stored snapshot $\mathbf{x}(t_k)$}
     	\If {a future snapshot exists at or near \(t_k + H\)} 
         	\State Find the closest future snapshot to \(t_k + H\)
		\State Assign future fragmentation label \(y(t_k)\)
		\State Write \([\mathbf{x}(t_k), y(t_k)]\) to the CSV dataset
        \EndIf
     \EndFor
     \State Destroy the ns-3 simulation state
 \EndFor
 \State Close the CSV file
 \State \Return $\mathcal{D}$
 \end{algorithmic}
 \end{algorithm}
 
 Snapshots for which no future observation is available at or near \(t_k + H\) are not written to the final dataset. This prevents the generation of incomplete samples and ensures that each exported observation is associated with a valid future fragmentation label.


\section{Simulation Scenario and Data Collection Profiles}\label{sect:s4}

 This section describes the simulation environment used to generate the UAV-assisted VANET dataset. It specifies the highway mobility model, the accident-induced slowdown mechanism, the UAV support placement strategy, the network and communication configuration, and the scenario profiles used to diversify the generated samples. These elements define the controlled experimental conditions under which mobility, topology, communication, and UAV coverage features are collected.

 \subsection{Highway Mobility Scenario}
 \label{subsec:highway_mobility}

 The simulated environment represents a two-lane highway in which vehicles move in opposite directions. The first lane is located at $y_1 = 100$ m and the second lane at $y_2 = 160$ m. Vehicles in the first lane move in the positive $x$-direction, whereas vehicles in the second lane move in the negative $x$-direction. The road length is configurable and denoted by $L$.

 Let $N_g$ be the number of ground vehicles. The vehicle set is divided approximately into two groups. For vehicles in lane 1, the initial position is defined as:
 \begin{equation}
 \mathbf{p}_i(0) = \left(100 + i s_1,\; y_1,\; 0\right),
 \label{eq:lane1_position}
 \end{equation}
 where $s_1$ is the initial spacing between consecutive vehicles. For vehicles in lane 2, the initial position is:
 \begin{equation}
 \mathbf{p}_i(0) = \left(L - 100 - i s_2,\; y_2,\; 0\right),
 \label{eq:lane2_position}
 \end{equation}
 where $s_2$ denotes the spacing in the opposite lane.

 Vehicle mobility is implemented using a constant-velocity model. Nominal speeds are assigned with controlled heterogeneity:
 \begin{equation}
 v_i =
 \begin{cases}
 20 + (i \bmod 5), & \text{if vehicle } i \text{ belongs to lane 1},\\
 18 + (i \bmod 4), & \text{if vehicle } i \text{ belongs to lane 2}.
 \end{cases}
 \label{eq:nominal_speed}
 \end{equation}

 To maintain traffic continuity during the full simulation, vehicles are wrapped around the road when they move beyond the simulated road boundaries. This mechanism keeps the highway populated without requiring dynamic insertion of new vehicles during an episode.

 \subsection{Accident and Slowdown Modeling}
 \label{subsec:accident_modeling}

 A local accident zone is optionally introduced to create a realistic disturbance in vehicle mobility. The accident zone is defined by its center position $x_a$, half-width $w_a$, start time $t_a^{\mathrm{start}}$, end time $t_a^{\mathrm{end}}$, and slowdown speed $v_a$. A vehicle is considered inside the accident zone when:
 \begin{equation}
 |x_i(t) - x_a| \leq w_a.
 \label{eq:accident_zone_condition}
 \end{equation}

 During the active accident interval, the speed of a vehicle inside the affected area is reduced to:
 \begin{equation}
 v_i(t) =
 \begin{cases}
 v_a, & t_a^{\mathrm{start}} \leq t \leq t_a^{\mathrm{end}}
 \text{ and } |x_i(t)-x_a| \leq w_a,\\
 v_i^{\mathrm{nominal}}, & \text{otherwise}.
 \end{cases}
 \label{eq:accident_speed}
 \end{equation}

 This mechanism creates local compression and expansion waves in the traffic stream. Vehicles entering the slowdown zone reduce their speed, while vehicles outside the zone continue moving at nominal speed. As a result, the spatial distribution of vehicles becomes uneven, which may increase the largest longitudinal gap, alter the number of connected components, and increase the probability of VANET fragmentation.

 \subsection{UAV Support Placement and Mobility}
 \label{subsec:uav_placement}

 UAVs are deployed above the road and act as temporary aerial support nodes. In the default configuration, two UAVs are initially positioned at approximately $0.35L$ and $0.65L$ along the road axis:
 \begin{equation}
 \mathbf{p}_{u_1}(0)
 =
 \left(0.35L,\; \frac{y_1+y_2}{2},\; h_u\right),
 \label{eq:uav1_position}
 \end{equation}
 \begin{equation}
 \mathbf{p}_{u_2}(0)
 =
 \left(0.65L,\; \frac{y_1+y_2}{2},\; h_u\right),
 \label{eq:uav2_position}
 \end{equation}
 where $h_u$ denotes the UAV altitude. In the baseline policy, UAVs remain static. In the proactive policy, UAVs move toward regions where the current risk score indicates increasing fragmentation.

 For a UAV $u$ and a target point $\mathbf{q}$, the UAV velocity is computed as:
 \begin{equation}
 \mathbf{v}_u(t)
 =
 \min
 \left(
 \frac{\gamma \|\mathbf{q}-\mathbf{p}_u(t)\|}{\Delta t},
 v_u^{\max}
 \right)
 \frac{\mathbf{q}-\mathbf{p}_u(t)}
 {\|\mathbf{q}-\mathbf{p}_u(t)\|},
 \label{eq:uav_velocity}
 \end{equation}
 where $\gamma$ is a risk-dependent control gain, $\Delta t$ is the sampling period, and $v_u^{\max}$ is the maximum UAV speed.

 The target point of the proactive policy is selected according to the current fragmentation condition:
\begin{equation}
\mathbf{q}(t)=
\begin{cases}
\mathbf{p}_u(t), & \text{if the risk is stable},\\[2pt]
\mathbf{c}_g(t), & \text{if the risk is moderate},\\[2pt]
\dfrac{\mathbf{c}_1(t)+\mathbf{c}_2(t)}{2}, &
\text{if the risk is severe or imminent}.
\end{cases}
\label{eq:uav_target}
\end{equation}
 where $\mathbf{c}_g(t)$ is the ground vehicle centroid, and $\mathbf{c}_1(t)$ and $\mathbf{c}_2(t)$ are the centroids of the two largest connected components.

 \subsection{Network and Communication Configuration}
 \label{subsec:network_configuration}

 The communication layer is configured using IEEE 802.11p in ad hoc mode. The wireless channel uses a constant-speed propagation delay model and a range-based propagation loss model parameterized by $R_{\mathrm{eff}}$. The network stack uses AODV routing and IPv4 addressing.

 UDP traffic is generated from vehicles toward UAV sink nodes. Two traffic categories are modeled. The first category corresponds to base traffic, representing regular vehicular communication. The second category corresponds to burst traffic, representing temporary communication stress caused by events such as alerts, congestion, or local disturbances.

 For each traffic flow, packet size and data rate are randomized within profile-specific ranges:
 \begin{equation}
 S_p \sim \mathcal{U}(S_{\min}, S_{\max}),
 \label{eq:packet_size}
 \end{equation}
 \begin{equation}
 r_p \sim \mathcal{U}(r_{\min}, r_{\max}),
 \label{eq:data_rate}
 \end{equation}
 where $S_p$ is the packet size and $r_p$ is the data rate. Base traffic typically uses lower data rates, whereas burst traffic uses higher rates and shorter activation intervals. This design enables the dataset to capture both normal communication and stressed communication windows.

 \subsection{Data Collection Profiles}
 \label{subsec:data_profiles}

 To diversify the generated dataset, the simulator supports five scenario profiles. When scenario balancing is enabled, profiles are cycled across episodes to avoid over-representing a single mobility or traffic condition.
 
 Table~\ref{tab:data_profiles} summarizes the five scenario profiles supported by the generator, including their main purpose and typical configuration effects.

\begin{table}[H]
 \centering
 \caption{Data collection profiles supported by the proposed generator.\label{tab:data_profiles}}
 \begin{tabular*}{\textwidth}{@{\extracolsep\fill}p{0.08\textwidth}p{0.18\textwidth}p{0.31\textwidth}p{0.34\textwidth}}
 \toprule
 \textbf{Profile ID} &
 \textbf{Profile name} &
 \textbf{Main purpose} &
 \textbf{Typical configuration effect} \\
 \midrule
 0 &
 \texttt{free\_flow} &
 Represents normal highway traffic. &
 Moderate density, no accident, and limited burst traffic stress. \\

 1 &
 \texttt{local\_accident} &
 Models localized mobility disruption. &
 Accident zone, local slowdown, and moderate-to-high traffic stress. \\

 2 &
 \texttt{sparse\_stretched} &
 Increases fragmentation probability. &
 Fewer vehicles, longer road length, and lower effective communication range. \\

 3 &
 \texttt{dense\_burst} &
 Tests communication stress under dense mobility. &
 Higher vehicle density and stronger burst traffic. \\

 4 &
 \texttt{mixed\_stress} &
 Combines multiple adverse factors. &
 Accident, lower range, burst traffic, and variable topology. \\
 \bottomrule
 \end{tabular*}
 \end{table}

 The \texttt{free\_flow} profile is intended to generate mostly stable samples, although temporary risk may still arise due to mobility and UAV placement. The \texttt{local\_accident} profile introduces localized slowdowns that can create uneven vehicle distributions. The \texttt{sparse\_stretched} profile is designed to produce higher fragmentation risk by increasing road length and reducing density. The \texttt{dense\_burst} profile evaluates whether communication stress can degrade packet delivery even when the topology is relatively dense. Finally, the \texttt{mixed\_stress} profile combines mobility, topology, and traffic stressors to generate challenging samples for supervised learning.


\section{Feature Extraction and Future Labeling Methodology}
 \label{sect:s5}

 This section formalizes the transformation of raw simulation states into machine-learning-ready samples. The proposed methodology combines periodic sampling, mobility descriptors, graph-based topology metrics, UAV coverage indicators, and communication-window statistics. It also defines a future-oriented labeling strategy that assigns each current network state to a short-term fragmentation class observed after a configurable prediction horizon.

 \subsection{Sampling Process}
 \label{subsec:sampling_process}

 The simulation is sampled periodically after an initial warm-up time. Let:
 \begin{equation}
 t_k = t_0 + k\Delta t,
 \label{eq:sampling_time}
 \end{equation}
 where $t_0$ is the sampling start time, $\Delta t$ is the sampling period, and $k$ is the sampling index. At each sampling time, the generator extracts a feature vector from the current simulation state. The learning problem is formulated as:
 \begin{equation}
 f_{\theta}: \mathbf{x}_{t_k} \rightarrow y_{t_k},
 \label{eq:learning_mapping}
 \end{equation}
 where $\mathbf{x}_{t_k}$ is the feature vector extracted at time $t_k$, and $y_{t_k}$ is the fragmentation class observed at approximately $t_k + H$.

 \subsection{Mobility Features}
 \label{subsec:mobility_features}

 The first feature group describes the spatial and kinematic state of the vehicles. The speed of vehicle $i$ at time $t$ is:
 \begin{equation}
 v_i(t)
 =
 \sqrt{
 v_{x,i}(t)^2+
 v_{y,i}(t)^2+
 v_{z,i}(t)^2
 }.
 \label{eq:vehicle_speed}
 \end{equation}

 The average vehicle speed is:
 \begin{equation}
 \bar{v}(t)
 =
 \frac{1}{N_g}
 \sum_{i=1}^{N_g} v_i(t),
 \label{eq:avg_speed}
 \end{equation}
 and the speed variance is:
 \begin{equation}
 \sigma_v^2(t)
 =
 \frac{1}{N_g}
 \sum_{i=1}^{N_g}
 \left(v_i(t)-\bar{v}(t)\right)^2.
 \label{eq:speed_variance}
 \end{equation}

 The average inter-vehicle distance is computed over all vehicle pairs:
 \begin{equation}
 \bar{d}_{vv}(t)
 =
 \frac{2}{N_g(N_g-1)}
 \sum_{i=1}^{N_g-1}
 \sum_{j=i+1}^{N_g}
 d_{ij}^{2D}(t),
 \label{eq:avg_inter_vehicle_distance}
 \end{equation}
 where
 \begin{equation}
 d_{ij}^{2D}(t)
 =
 \sqrt{
 (x_i(t)-x_j(t))^2+
 (y_i(t)-y_j(t))^2
 }.
 \label{eq:2d_distance}
 \end{equation}

 The largest longitudinal gap is obtained by sorting vehicle positions along the $x$-axis:
 \begin{equation}
 x_{(1)}(t) \leq x_{(2)}(t) \leq \cdots \leq x_{(N_g)}(t).
 \label{eq:sorted_positions}
 \end{equation}
 The largest gap is then:
 \begin{equation}
 g_{\max}(t)
 =
 \max
 \left(
 \max_{2 \leq i \leq N_g}
 \left[x_{(i)}(t)-x_{(i-1)}(t)\right],
 L - x_{(N_g)}(t) + x_{(1)}(t)
 \right).
 \label{eq:largest_gap}
 \end{equation}

 The fraction of slow vehicles is computed as:
 \begin{equation}
 \rho_{\mathrm{slow}}(t)
 =
 \frac{1}{N_g}
 \sum_{i=1}^{N_g}
 \indicator
 \left[
 v_i(t) \leq v_{\mathrm{slow}}
 \right],
 \label{eq:fraction_slow}
 \end{equation}
 where $v_{\mathrm{slow}}$ is a slowdown threshold defined by:
 \begin{equation}
 v_{\mathrm{slow}} =
 \max(8,\; v_a + 2).
 \label{eq:slow_threshold}
 \end{equation}

 \subsection{Topology Features}
 \label{subsec:topology_features}

 Topology features are computed from the connectivity graph $G(t)$. The degree of node $i$ is:
 \begin{equation}
 k_i(t) = |\mathcal{N}_i(t)|,
 \label{eq:node_degree}
 \end{equation}
 where $\mathcal{N}_i(t)$ is the set of one-hop neighbors of node $i$. The average degree is:
 \begin{equation}
 \bar{k}(t)
 =
 \frac{1}{N}
 \sum_{i=1}^{N} k_i(t),
 \label{eq:avg_degree}
 \end{equation}
 where $N=N_g+N_u$ is the total number of ground and aerial nodes. The degree variance is:
 \begin{equation}
 \sigma_k^2(t)
 =
 \frac{1}{N}
 \sum_{i=1}^{N}
 \left(k_i(t)-\bar{k}(t)\right)^2.
 \label{eq:degree_variance}
 \end{equation}

 Connected components are identified using breadth-first search over the adjacency list. Let:
 \begin{equation}
 \mathcal{C}(t)
 =
 \{C_1(t), C_2(t), \ldots, C_K(t)\}
 \label{eq:components}
 \end{equation}
 be the set of connected components, where $K(t)$ is the number of components. The largest connected component size is:
 \begin{equation}
 LCC(t)
 =
 \max_{1 \leq k \leq K(t)}
 |C_k(t)|.
 \label{eq:lcc}
 \end{equation}
 The normalized largest connected component ratio is:
 \begin{equation}
 NLCC(t)
 =
 \frac{LCC(t)}{N}.
 \label{eq:nlcc}
 \end{equation}

 A high value of $NLCC(t)$ indicates that most nodes belong to a single connected component, whereas a low value indicates increasing fragmentation.

 \subsection{UAV Coverage Features}
 \label{subsec:uav_coverage_features}

 UAV coverage features measure how well aerial support nodes support the ground network. The ground vehicle centroid is:
 \begin{equation}
 \mathbf{c}_g(t)
 =
 \frac{1}{N_g}
 \sum_{i=1}^{N_g}
 \mathbf{p}_i(t).
 \label{eq:ground_centroid}
 \end{equation}

 The average UAV distance to the ground centroid is:
 \begin{equation}
 \bar{d}_{u,c}(t)
 =
 \frac{1}{N_u}
 \sum_{u=1}^{N_u}
 d^{2D}(\mathbf{p}_u(t), \mathbf{c}_g(t)).
 \label{eq:uav_centroid_distance}
 \end{equation}

 A vehicle $i$ is considered covered by UAV $u$ if:
 \begin{equation}
 d_{iu}^{3D}(t) \leq R_{\mathrm{eff}}.
 \label{eq:uav_vehicle_coverage}
 \end{equation}

 The number of uniquely covered vehicles is:
 \begin{equation}
 N_{\mathrm{cov}}(t)
 =
 \sum_{i=1}^{N_g}
 \indicator
 \left[
 \exists u \in V_u(t):
 d_{iu}^{3D}(t) \leq R_{\mathrm{eff}}
 \right].
 \label{eq:num_covered_vehicles}
 \end{equation}

 The UAV coverage ratio is then:
 \begin{equation}
 \rho_{\mathrm{uav}}(t)
 =
 \frac{N_{\mathrm{cov}}(t)}{N_g}.
 \label{eq:uav_coverage_ratio}
 \end{equation}

 \subsection{Traffic-Window Features}
 \label{subsec:traffic_features}

 The generator maintains cumulative counters for transmitted packets, received packets, transmitted bytes, and received bytes. At each sampling time, window-level traffic features are computed using counter differences between the current and previous sampling instants.

 Let $P_{tx}(t_k)$ and $P_{rx}(t_k)$ be the cumulative transmitted and received packet counters. The number of packets sent during the current window is:
 \begin{equation}
 \Delta P_{tx}(t_k)
 =
 P_{tx}(t_k)-P_{tx}(t_{k-1}),
 \label{eq:delta_tx_packets}
 \end{equation}
 and the number of packets received is:
 \begin{equation}
 \Delta P_{rx}(t_k)
 =
 P_{rx}(t_k)-P_{rx}(t_{k-1}).
 \label{eq:delta_rx_packets}
 \end{equation}

 The packet delivery ratio over the current window is:
 \begin{equation}
 PDR(t_k)
 =
 \frac{\Delta P_{rx}(t_k)}
 {\Delta P_{tx}(t_k)},
 \quad
 \Delta P_{tx}(t_k) > 0.
 \label{eq:pdr_window}
 \end{equation}
 
 When no packet is transmitted during a sampling window, i.e., \(\Delta P_{tx}(t_k)=0\), the implementation retains the last valid PDR value. The PDR is initialized to 1.0 before the first valid traffic window.

 The throughput in Kbps is:
 \begin{equation}
 T_{\mathrm{kbps}}(t_k)
 =
 \frac{8 \cdot \Delta B_{rx}(t_k)}
 {1000 \cdot \Delta t},
 \label{eq:throughput_window}
 \end{equation}
 where $\Delta B_{rx}(t_k)$ is the number of bytes received during the sampling window.

 \subsection{Current Risk Score}
 \label{subsec:current_risk_score}

 Although the supervised target is computed from a future state, the generator also computes a current heuristic risk score used for logging and proactive UAV control. The score combines topology degradation, communication quality, mobility disturbance, longitudinal separation, and UAV coverage.

 The component risk is:
 \begin{equation}
 R_{\mathrm{comp}}(t) = 1 - NLCC(t).
 \label{eq:component_risk}
 \end{equation}

 The degree risk is:
 \begin{equation}
 R_{\mathrm{deg}}(t)
 =
 \clip
 \left(
 1 - \frac{\bar{k}(t)}{4},
 0,
 1
 \right).
 \label{eq:degree_risk}
 \end{equation}

 The PDR risk is:
 \begin{equation}
 R_{\mathrm{pdr}}(t)
 =
 1 - PDR(t).
 \label{eq:pdr_risk}
 \end{equation}

 The slow-vehicle risk is:
 \begin{equation}
 R_{\mathrm{slow}}(t)
 =
 \clip
 \left(
 \rho_{\mathrm{slow}}(t),
 0,
 1
 \right).
 \label{eq:slow_risk}
 \end{equation}

 The gap risk is:
 \begin{equation}
 R_{\mathrm{gap}}(t)
 =
 \clip
 \left(
 \frac{g_{\max}(t)}{350},
 0,
 1
 \right).
 \label{eq:gap_risk}
 \end{equation}

 The component penalty is defined as:
 \begin{equation}
 R_{\mathrm{pen}}(t)
 =
 \begin{cases}
 0, & K(t) \leq 1,\\
 0.45, & K(t)=2,\\
 0.75, & K(t)=3,\\
 1.0, & K(t)\geq 4.
 \end{cases}
 \label{eq:component_penalty}
 \end{equation}

 The final current risk score is:
 \begin{align}
 R(t) =
 &
 0.28R_{\mathrm{comp}}(t)
 +
 0.10R_{\mathrm{deg}}(t)
 +
 0.18R_{\mathrm{pdr}}(t)
 \nonumber\\
 &
 +
 0.16R_{\mathrm{pen}}(t)
 +
 0.12R_{\mathrm{slow}}(t)
 +
 0.10R_{\mathrm{gap}}(t)
 +
 0.06\left(1-\rho_{\mathrm{uav}}(t)\right).
 \label{eq:current_risk_score}
 \end{align}

 The score is clipped to the interval $[0,1]$ and mapped to a current risk class:
 \begin{equation}
 \mathrm{Risk}(t)=
 \begin{cases}
 3, & NLCC(t)<0.35 \vee K(t)\geq4 \vee R(t)\geq0.75,\\
 2, & NLCC(t)<0.55 \vee K(t)\geq3 \vee R(t)\geq0.50,\\
 1, & NLCC(t)<0.80 \vee K(t)=2 \vee R(t)\geq0.22,\\
 0, & \text{otherwise}.
 \end{cases}
 \label{eq:current_risk_class}
 \end{equation}

 The classes correspond respectively to stable topology, moderate fragmentation risk, severe fragmentation risk, and imminent fragmentation.
 
 In the implementation, the risk classification additionally accounts for severe packet delivery degradation. Very low PDR values are treated as supplementary indicators of connectivity degradation and may increase the assigned risk level when combined with fragmented network topologies.

 \subsection{Future Label Construction}
 \label{subsec:future_labeling}

 The final supervised learning target is generated from the future topology rather than the current heuristic score. For a sample extracted at $t_k$, the generator selects the stored snapshot closest to:
 \begin{equation}
 t_k^{\mathrm{future}} = t_k + H.
 \label{eq:future_time}
 \end{equation}

 Let the future snapshot be denoted by $\mathbf{x}_{t_j}$, where:
 \begin{equation}
 j =
 \arg\min_{m>k}
 |t_m - (t_k + H)|.
 \label{eq:closest_future_snapshot}
 \end{equation}

 The future label is assigned using the future values of $NLCC(t_j)$ and $K(t_j)$:
 \begin{equation}
 y_{t_k}
 =
 \begin{cases}
 0, & NLCC(t_j)\geq0.80 \wedge K(t_j)=1,\\
 1, & NLCC(t_j)\geq0.60 \wedge K(t_j)\leq2,\\
 2, & NLCC(t_j)\geq0.35 \wedge K(t_j)\leq3,\\
 3, & \text{otherwise}.
 \end{cases}
 \label{eq:future_label}
 \end{equation}
 
 Algorithm~\ref{alg:future_labeling} summarizes the feature extraction and future-labeling procedure used to transform stored simulation snapshots into labeled supervised learning samples.

 This strategy creates a predictive dataset rather than a purely descriptive one. A future supervised learning model can then be trained to infer a future fragmentation condition from the current mobility, topology, UAV coverage, and communication state.

 \begin{algorithm}[H]
 \caption{Feature extraction and future labeling}
 \label{alg:future_labeling}
 \begin{algorithmic}[1]
 \Require Stored snapshots $\mathcal{S} = \{\mathbf{x}(t_1), \mathbf{x}(t_2), \ldots, \mathbf{x}(t_M)\}$, prediction horizon $H$
 \Ensure Labeled dataset $\mathcal{D}$
 \For{each snapshot $\mathbf{x}(t_i)$ in $\mathcal{S}$}
     \State $t_{\mathrm{target}} \gets t_i + H$
     \State $j \gets \arg\min_{m>i} |t_m - t_{\mathrm{target}}|$
     \If{no valid future snapshot exists}
         \State \textbf{continue}
     \EndIf
     \State $\mathbf{x}_{\mathrm{future}} \gets \mathbf{x}(t_j)$
     \If{$NLCC(t_j) \geq 0.80$ and $K(t_j)=1$}
         \State $y \gets 0$
     \ElsIf{$NLCC(t_j) \geq 0.60$ and $K(t_j)\leq2$}
         \State $y \gets 1$
     \ElsIf{$NLCC(t_j) \geq 0.35$ and $K(t_j)\leq3$}
         \State $y \gets 2$
     \Else
         \State $y \gets 3$
     \EndIf
     \State Write $\left[\mathbf{x}(t_i), y\right]$ to $\mathcal{D}$
 \EndFor
 \State \Return $\mathcal{D}$
 \end{algorithmic}
 \end{algorithm}


\section{Reproducibility, Implementation, and Dataset Generation Protocol}\label{sect:s6}

This section presents the implementation and reproducibility mechanisms of the proposed generator. Beyond the conceptual design, reproducibility depends on controlled software dependencies, deterministic randomization, explicit scenario configuration, standardized output formats, and a repeatable dataset generation protocol. These elements are described to facilitate independent replication, extension, and comparative evaluation of the generated datasets.

 \subsection{Implementation Environment}
 \label{subsec:implementation_environment}

 The generator is implemented as a standalone ns-3-based C++ project. The implementation uses the ns-3 modules required for node creation, mobility modeling, wireless communication, Internet stack installation, routing, application-level traffic generation, flow monitoring, and optional visualization. The implementation is intentionally self-contained: vehicle mobility, UAV mobility, traffic generation, feature extraction, labeling, CSV writing, and optional visualization are handled within the same generation pipeline. This design reduces external dependencies during dataset generation and supports controlled reproducibility across simulation runs.

 \subsection{Reproducibility Controls}
 \label{subsec:reproducibility_controls}

 Several design choices are used to improve reproducibility. First, all randomized episode parameters are controlled by a configurable random seed:
 \begin{equation}
 s \in \mathbb{N}.
 \label{eq:random_seed}
 \end{equation}
 Using the same seed and the same command-line arguments produces the same sequence of scenario profiles, mobility parameters, traffic rates, packet sizes, accident configurations, and UAV policies.

 Second, scenario balancing can be enabled so that the generator cycles deterministically through the five predefined profiles:
 \begin{equation}
 p_e = e \bmod 5,
 \label{eq:scenario_balancing}
 \end{equation}
 where $e$ is the episode index. This prevents the dataset from being dominated by a single scenario type.

 Third, UAV policies can be alternated across episodes:
 \begin{equation}
 \pi_e =
 \begin{cases}
 \text{baseline}, & e \bmod 2 = 0,\\
 \text{proactive}, & e \bmod 2 = 1.
 \end{cases}
 \label{eq:policy_cycling}
 \end{equation}

 This policy cycling enables direct comparison between static UAV support nodes and proactive UAV repositioning under comparable episode distributions.

 \subsection{Dataset Output Structure}
 \label{subsec:dataset_output_structure}

 Each generated row contains episode metadata, scenario parameters, mobility features, topology features, UAV coverage features, traffic-window features, contextual indicators, current risk information, and the future label.
 
 Table~\ref{tab:feature_groups} summarizes the main feature groups exported by the generator, representative CSV columns, and their interpretation.

\begin{table}[!htbp]
 \centering
 \caption{Main feature groups in the generated dataset.\label{tab:feature_groups}}
 \begin{tabular*}{\textwidth}{@{\extracolsep\fill}p{0.20\textwidth}p{0.38\textwidth}p{0.34\textwidth}}
 \toprule
 \textbf{Feature group} &
 \textbf{Representative columns} &
 \textbf{Interpretation} \\
 \midrule
 Episode metadata &
 \texttt{episode\_id}, \texttt{scenario\_profile}, \texttt{scenario\_profile\_name}, \texttt{policy\_proactive\_uav} &
 Identifies the episode, scenario family, and UAV policy. \\

 Scenario parameters &
 \texttt{num\_vehicles}, \texttt{num\_uavs}, \texttt{effective\_range}, \texttt{road\_length}, \texttt{base\_sources}, \texttt{burst\_sources} &
 Describes the structural configuration of the episode. \\

 Accident context &
 \texttt{accident\_enabled}, \texttt{accident\_start}, \texttt{accident\_end}, \texttt{accident\_x}, \texttt{accident\_active\_now} &
 Indicates whether and when a local slowdown affects mobility. \\

 Mobility features &
 \texttt{avg\_vehicle\_speed}, \texttt{speed\_variance}, \texttt{avg\_inter\_vehicle\_distance}, \texttt{largest\_longitudinal\_gap}, \texttt{fraction\_slow\_vehicles} &
 Captures vehicle distribution and mobility disturbances. \\

 Topology features &
 \texttt{num\_components}, \texttt{largest\_component\_size}, \texttt{average\_degree}, \texttt{degree\_variance}, \texttt{normalized\_largest\_component} &
 Captures graph-level connectivity and fragmentation. \\

 UAV coverage features &
 \texttt{avg\_uav\_distance\_to\_ground\_centroid}, \texttt{num\_vehicles\_covered\_by\_uav}, \texttt{uav\_coverage\_ratio} &
 Measures aerial support coverage. \\

 Traffic-window features &
 \texttt{packets\_sent\_window}, \texttt{packets\_received\_window}, \texttt{tx\_bytes\_window}, \texttt{rx\_bytes\_window}, \texttt{pdr\_window}, \texttt{throughput\_kbps\_window} &
 Captures communication performance over the sampling window. \\

 Risk and label &
 \texttt{current\_risk\_score}, \texttt{current\_risk\_class}, \texttt{future\_label} &
 Provides current heuristic risk and the supervised future target. \\
 \bottomrule
 \end{tabular*}
 \end{table}
 
\vspace{-0.8em}

 The inclusion of metadata is important because it enables stratified evaluation. For instance, a model can be trained on all profiles and then evaluated separately on sparse-stretched or mixed-stress scenarios to assess generalization under difficult conditions.
 
 \subsection{Dataset Generation Protocol}
 \label{subsec:generation_protocol}

 The recommended dataset generation protocol consists of four stages. The first stage is visual validation, where a short simulation is executed with visualization enabled to verify vehicle placement, UAV positions, traffic flow activation, and general scenario behavior. The second stage is a small randomized run, where a limited number of episodes is generated to verify the CSV structure, label distribution, traffic counters, and scenario diversity. The third stage is large-scale dataset generation, where visualization is disabled and a large number of randomized and balanced episodes is executed. The fourth stage is multi-seed aggregation, where multiple datasets generated with different seeds are merged to improve diversity and reduce dependence on a single random sequence.
 
 Algorithm~\ref{alg:multi_seed_generation} summarizes the recommended multi-seed dataset generation protocol.

{\scriptsize
 \begin{algorithm}[H]
 \caption{Reproducible multi-seed dataset generation protocol}
 \label{alg:multi_seed_generation}
 \begin{algorithmic}[1]
 \Require Seed set $\mathcal{S} = \{s_1, s_2, \ldots, s_m\}$, number of episodes per seed $N_e$
 \Ensure Merged dataset $\mathcal{D}_{\mathrm{merged}}$
 \State $\mathcal{D}_{\mathrm{merged}} \gets \emptyset$
 \For{each seed $s$ in $\mathcal{S}$}
     \State Configure the generator with \texttt{rngSeed} $=s$
     \State Enable randomized episodes
     \State Enable scenario profile balancing
     \State Enable UAV policy cycling
     \State Disable visualization for scalability
     \State Run $N_e$ simulation episodes
     \State Export CSV file $\mathcal{D}_s$
     \State Append $\mathcal{D}_s$ to $\mathcal{D}_{\mathrm{merged}}$
 \EndFor
 \State Remove duplicate rows if necessary
 \State Check missing values and invalid feature ranges
 \State Report class distribution by scenario profile and UAV policy
 \State Save $\mathcal{D}_{\mathrm{merged}}$
 \State \Return $\mathcal{D}_{\mathrm{merged}}$
 \end{algorithmic}
 \end{algorithm}
 }
 
 \subsection{Illustrative Generation Command}
 \label{subsec:illustrative_generation_command}

 To demonstrate the practical execution of the proposed generator, an illustrative dataset instance was generated using a balanced and randomized configuration. The run used 20 simulation episodes, randomized episode parameters, balanced scenario profiles, alternating UAV policies, enabled traffic generation, a sampling period of 0.5~s, a sampling start time of 4.0~s, a prediction horizon of 5.0~s, and a total simulation duration of 120~s. The following command was used:

\begin{samepage}
{\scriptsize
\begin{verbatim}
./build/bin/ns3.46.1-uav_vanet_ml_runner-debug \
  --numEpisodes=20 \
  --randomizeEpisodes=true \
  --balanceScenarioProfiles=true \
  --cyclePolicyAcrossEpisodes=true \
  --enableNetAnim=false \
  --enableTraffic=true \
  --samplePeriod=0.5 \
  --samplingStartTime=4.0 \
  --predictionHorizon=5.0 \
  --simTime=120 \
  --eventGuardTime=0.001 \
  --csvFile=scratch/uav-vanet-ml/data/output.csv \
  --rngSeed=42
\end{verbatim}
}
\end{samepage}

 This configuration was selected to verify the ability of the generator to produce a balanced multi-scenario dataset while maintaining reproducibility through a fixed random seed. The generated CSV file is used in Section~\ref{sect:s7} to characterize the dataset structure, scenario balance, label distribution, and feature ranges. The objective of this analysis is not to report machine learning performance, but to validate that the generator produces coherent and reusable data for subsequent predictive modeling studies.

 \subsection{Recommended Machine Learning Use}
 \label{subsec:recommended_ml_use}

 Although the present article focuses on dataset generation rather than predictive model optimization, the generated data are structured to support future supervised learning experiments. The input matrix is:

 \begin{equation}
 \mathbf{X}
 =
 [
 \mathbf{x}_{t_1},
 \mathbf{x}_{t_2},
 \ldots,
 \mathbf{x}_{t_M}
 ]^T,
 \label{eq:input_matrix}
 \end{equation}
 and the target vector is:
 \begin{equation}
 \mathbf{y}
 =
 [
 y_{t_1},
 y_{t_2},
 \ldots,
 y_{t_M}
 ]^T.
 \label{eq:target_vector}
 \end{equation}

 The recommended evaluation protocol includes accuracy, macro F1-score, weighted F1-score, balanced accuracy, class-wise precision, class-wise recall, and confusion matrices. Macro F1-score is particularly important because severe and imminent fragmentation samples may be less frequent than stable samples, depending on the selected scenario mix.

A suitable experimental design is to compare the baseline UAV policy against the proactive UAV policy under the same scenario distribution and evaluation protocol.

 This comparison can be performed in two complementary ways. First, the generated labels can be analyzed to determine whether proactive UAV movement reduces severe and imminent future fragmentation. Second, machine learning models can be trained to predict fragmentation risk and evaluated across policies and scenario profiles.


\section{Generated Dataset Characterization, Discussion, and Conclusion}
 \label{sect:s7}

 This section characterizes the dataset generated by the proposed UAV-assisted VANET framework using the illustrative run described in Section~\ref{subsec:illustrative_generation_command}. The purpose of this characterization is to verify the behavior of the generator, assess scenario and policy balance, inspect the distribution of future fragmentation labels, and confirm that the extracted features remain within coherent ranges. Since the present article focuses on the generator rather than on machine learning model optimization, the analysis remains descriptive and does not report classifier performance.

 \subsection{Illustrative Dataset Instance}
 \label{subsec:illustrative_dataset_instance}

 The illustrative execution produced a dataset containing 4,620 samples collected over 20 independent simulation episodes. Sampling started at 4.0~s and continued until 119.0~s with a sampling period of 0.5~s. Each episode therefore contributed a sequence of time-indexed observations describing the current mobility state, connectivity topology, UAV coverage, communication traffic window, current heuristic risk score, and future fragmentation label.

 Table~\ref{tab:generated_dataset_summary} summarizes the main characteristics of the generated dataset instance. The dataset contains five scenario profiles and two UAV policy configurations. The scenario profile balancing mechanism produced the same number of samples for each profile, while the UAV policy cycling mechanism produced an equal number of samples for the baseline and proactive UAV policies. This confirms that the generator can produce controlled and balanced simulation traces suitable for downstream comparative analysis.

\begin{table}[H]
 \centering
 \caption{Summary of the illustrative generated dataset instance.}
 \label{tab:generated_dataset_summary}
 \begin{tabular*}{\textwidth}{@{\extracolsep\fill}ll}
 \toprule
 \textbf{Property} & \textbf{Value} \\
 \midrule
 Number of episodes & 20 \\
 Number of samples & 4,620 \\
 Sampling period & 0.5~s \\
 Sampling start time & 4.0~s \\
 Last sampled time & 119.0~s \\
 Prediction horizon & 5.0~s \\
 Scenario profiles & free\_flow, local\_accident, sparse\_stretched, dense\_burst, mixed\_stress \\
 Samples per scenario profile & 924 \\
 Baseline UAV policy samples & 2,310 \\
 Proactive UAV policy samples & 2,310 \\
 Traffic generation & Enabled \\
 NetAnim visualization & Disabled \\
 Random seed & 42 \\
 \bottomrule
 \end{tabular*}
 \end{table}
 
 \subsection{Scenario and Policy Balance}
 \label{subsec:scenario_policy_balance}

 Scenario balance is important because a dataset dominated by a single traffic condition may bias subsequent learning tasks and reduce the ability to evaluate model behavior under heterogeneous network states. In the generated dataset, each of the five scenario profiles contributes exactly 924 samples, representing 20\% of the dataset. This confirms that the profile balancing mechanism works as intended.

 The UAV policy distribution is also balanced. The baseline UAV policy and the proactive UAV policy each contribute 2,310 samples, corresponding to 50\% of the generated observations. This balance enables controlled comparison between static aerial relay behavior and risk-aware UAV repositioning without conflating policy effects with differences in scenario frequency.
 
 Table~\ref{tab:scenario_profile_distribution} reports the distribution of samples across the five scenario profiles and confirms the effect of scenario balancing.

\begin{table}[H]
 \centering
 \caption{Scenario profile distribution in the generated dataset.}
 \label{tab:scenario_profile_distribution}
 \begin{tabular*}{\textwidth}{@{\extracolsep\fill}lcc}
 \toprule
 \textbf{Scenario profile} & \textbf{Number of samples} & \textbf{Share of dataset} \\
 \midrule
 free\_flow & 924 & 20.0\% \\
 local\_accident & 924 & 20.0\% \\
 sparse\_stretched & 924 & 20.0\% \\
 dense\_burst & 924 & 20.0\% \\
 mixed\_stress & 924 & 20.0\% \\
 \bottomrule
 \end{tabular*}
 \end{table}
 
 \subsection{Future Fragmentation Label Distribution}
 \label{subsec:future_label_distribution}

 The future fragmentation label is the supervised target produced by the generator. It is computed from the future network state after the configured prediction horizon rather than from the current heuristic risk score. Table~\ref{tab:future_label_distribution} reports the distribution of future labels in the illustrative dataset. The generated data contain all four fragmentation classes, ranging from stable connectivity to imminent fragmentation.

 The stable class represents 45.65\% of the samples, while the imminent fragmentation class represents 29.89\%. Moderate and severe fragmentation account for 9.24\% and 15.22\% of the dataset, respectively. This distribution is suitable for validating the generator because it confirms that the scenario profiles produce both connected and fragmented future states. At the same time, the distribution remains moderately imbalanced, which is realistic for risk prediction problems and should be considered in future machine learning experiments.

\begin{table}[H]
 \centering
 \caption{Future fragmentation label distribution in the generated dataset.}
 \label{tab:future_label_distribution}
 \begin{tabular*}{\textwidth}{@{\extracolsep\fill}cclcc}
 \toprule
 \textbf{Label} & \textbf{Risk level} & \textbf{Interpretation} & \textbf{Samples} & \textbf{Share} \\
 \midrule
 0 & Stable & Connected or weakly fragmented future topology & 2,109 & 45.65\% \\
 1 & Moderate & Early degradation or limited fragmentation & 427 & 9.24\% \\
 2 & Severe & Significant fragmentation risk & 703 & 15.22\% \\
 3 & Imminent & Strong future fragmentation condition & 1,381 & 29.89\% \\
 \bottomrule
 \end{tabular*}
 \end{table}
 
 \subsection{Scenario-Specific Label Behavior}
 \label{subsec:scenario_specific_label_behavior}

 The label distribution by scenario profile provides an additional sanity check on the generator. As shown in Table~\ref{tab:label_distribution_by_profile}, the free\_flow and dense\_burst profiles generate no imminent fragmentation samples in the illustrative run, which is consistent with their relatively stable or dense topology assumptions. By contrast, the sparse\_stretched profile produces a high proportion of imminent fragmentation samples, reflecting the effect of reduced vehicle density, longer road length, and lower effective communication range. The mixed\_stress profile also generates a large number of imminent fragmentation samples because it combines several adverse factors, including accident conditions, burst traffic, and variable topology.

 These results indicate that the scenario profiles influence the future fragmentation labels in a coherent manner. The generated labels therefore reflect the intended stress mechanisms embedded in the simulator rather than random or arbitrary class assignment.

\begin{table}[H]
 \centering
 \caption{Future label distribution by scenario profile.}
 \label{tab:label_distribution_by_profile}
 \begin{tabular*}{\textwidth}{@{\extracolsep\fill}lcccc}
 \toprule
 \textbf{Scenario profile} & \textbf{Label 0} & \textbf{Label 1} & \textbf{Label 2} & \textbf{Label 3} \\
 \midrule
 free\_flow & 618 & 64 & 242 & 0 \\
 local\_accident & 384 & 151 & 181 & 208 \\
 sparse\_stretched & 83 & 36 & 65 & 740 \\
 dense\_burst & 674 & 80 & 170 & 0 \\
 mixed\_stress & 350 & 96 & 45 & 433 \\
 \bottomrule
 \end{tabular*}
 \end{table}
 
 \subsection{Feature Range and Sanity Checks}
 \label{subsec:feature_range_sanity_checks}

 The generated dataset was also inspected to verify that the extracted features remain within coherent ranges. Table~\ref{tab:feature_range_summary} summarizes representative feature ranges observed in the illustrative dataset. The number of vehicles varies between 19 and 41, while the number of UAVs varies between 1 and 3, confirming that the randomized episode configuration introduces structural diversity. The effective communication range varies from approximately 196.82~m to 297.04~m, and the road length varies from approximately 2.47~km to 4.65~km.

 The communication-related indicators also show substantial variability. The packet delivery ratio ranges from 0 to 1, while the throughput reaches up to 3,617.89~Kbps. The UAV coverage ratio ranges from 0 to 0.862, indicating that the UAVs sometimes cover only a limited subset of vehicles and sometimes provide broader aerial support. These observations are consistent with the intended purpose of the generator, which is to expose learning models to heterogeneous mobility, topology, UAV coverage, and communication conditions.

\begin{table}[H]
 \centering
 \caption{Representative feature ranges in the illustrative generated dataset.}
 \label{tab:feature_range_summary}
 \begin{tabular*}{\textwidth}{@{\extracolsep\fill}lccc}
 \toprule
 \textbf{Feature} & \textbf{Minimum} & \textbf{Maximum} & \textbf{Mean} \\
 \midrule
 Number of vehicles & 19 & 41 & 30.85 \\
 Number of UAVs & 1 & 3 & 2.35 \\
 Effective communication range (m) & 196.82 & 297.04 & 240.54 \\
 Road length (m) & 2470.57 & 4651.38 & 3496.04 \\
 Average vehicle speed (m/s) & 10.47 & 20.75 & 20.10 \\
 Average inter-vehicle distance (m) & 479.22 & 1714.29 & 958.58 \\
 Largest longitudinal gap (m) & 121.82 & 2769.35 & 915.13 \\
 UAV coverage ratio & 0.00 & 0.862 & 0.314 \\
 Packet delivery ratio & 0.00 & 1.00 & 0.488 \\
 Throughput (Kbps) & 0.00 & 3617.89 & 901.57 \\
 Current risk score & 0.086 & 0.793 & 0.368 \\
 \bottomrule
 \end{tabular*}
 \end{table}
 
 \subsection{Descriptive Comparison of UAV Policies}
 \label{subsec:uav_policy_descriptive_comparison}

 The generated dataset also records whether each sample was produced under the baseline UAV policy or the proactive UAV policy. Table~\ref{tab:policy_label_distribution} reports the future label distribution by UAV policy. In the illustrative run, the proactive UAV policy produces fewer imminent fragmentation samples than the baseline policy. However, this observation should be interpreted as descriptive rather than conclusive, because the present dataset instance contains only 20 episodes and was generated primarily to validate the generator.

 A more robust statistical assessment of UAV policy impact would require multiple seeds, larger episode counts, and controlled hypothesis testing. Therefore, the role of this comparison in the present article is limited to demonstrating that the generator can record UAV policy metadata and produce policy-aware datasets suitable for future predictive and comparative studies.

\begin{table}[H]
 \centering
 \caption{Future label distribution by UAV policy in the illustrative generated dataset.}
 \label{tab:policy_label_distribution}
 \begin{tabular*}{\textwidth}{@{\extracolsep\fill}lcccc}
 \toprule
 \textbf{UAV policy} & \textbf{Label 0} & \textbf{Label 1} & \textbf{Label 2} & \textbf{Label 3} \\
 \midrule
 Baseline UAV policy & 999 & 207 & 347 & 757 \\
 Proactive UAV policy & 1110 & 220 & 356 & 624 \\
 \bottomrule
 \end{tabular*}
 \end{table}
 
 \subsection{Synthesis of Dataset Properties}
 \label{subsec:synthesis_dataset_properties}

 The descriptive analyses reported above confirm that the proposed generator produces a structured, balanced, and policy-aware dataset suitable for future predictive modeling studies. Rather than introducing a new experimental result, this subsection synthesizes the main dataset properties that support the scientific value of the generator.

 First, it is multi-scenario. The five predefined profiles expose the learning task to normal traffic, local accident disturbances, sparse stretched topologies, dense burst traffic, and mixed stress conditions. This diversity is important because fragmentation risk may emerge from different mechanisms. In sparse scenarios, fragmentation is primarily driven by vehicle separation and low effective communication range. In accident scenarios, risk may arise from localized slowdowns and heterogeneous spacing. In dense burst scenarios, topology may remain connected while packet delivery decreases due to traffic stress.

 Second, the dataset is multi-modal at the feature level. It does not rely exclusively on packet-level traffic indicators. Instead, it jointly captures mobility, topology, UAV coverage, and communication-window features. This is important because fragmentation is not purely a communication-layer phenomenon. A reduction in packet delivery may be caused by congestion, while a reduction in the largest connected component may be caused by spatial separation. Combining these feature families allows learning models to distinguish between traffic-induced degradation and topology-induced fragmentation.

 Third, the dataset is future-labeled. The target is computed from the state of the network after a prediction horizon rather than from the current state. This choice is central to the contribution of the work. It enables the dataset to support proactive decision-making, where a model can predict whether the current state is likely to evolve toward moderate, severe, or imminent fragmentation.

 Fourth, the dataset is policy-aware. Each row records whether the UAV policy is baseline or proactive. This makes it possible to evaluate whether proactive UAV movement changes the distribution of future fragmentation labels and whether prediction models generalize across UAV control strategies.

 Fifth, the dataset is reproducible and extensible. The random seed, scenario profile, mobility parameters, UAV configuration, traffic configuration, and labeling rules are explicitly controlled. New profiles, UAV policies, propagation models, or labeling thresholds can be incorporated without changing the entire architecture.

 \subsection{Discussion}
 \label{subsec:discussion}

 The proposed generator addresses an important methodological gap in VANET research. Many VANET studies evaluate routing, security, or communication performance under a limited number of manually configured scenarios. Although such studies are useful for protocol evaluation, they are less suitable for training and comparing machine learning models because they often produce scenario-specific results. In contrast, the proposed framework is designed to generate a broad and reusable dataset, where each row is a labeled observation that can be used for supervised learning.

 The use of graph-level indicators is particularly important. The number of connected components $K(t)$ and the normalized largest connected component ratio $NLCC(t)$ provide direct information about fragmentation. However, these indicators alone are not sufficient for proactive prediction. For example, a network may still be connected at time $t$, but a large longitudinal gap, increasing speed variance, low UAV coverage, or a sudden drop in packet delivery ratio may indicate that fragmentation is likely to occur at $t+H$. The proposed feature set therefore captures both the current topology and early warning signs of future degradation.

 The proactive UAV policy implemented in the generator is deliberately simple and interpretable. When risk is moderate, UAVs move toward the ground vehicle centroid. When risk becomes severe or imminent, UAVs move toward the region between the two largest connected components. This behavior reflects a practical relay strategy: UAVs should not merely follow the densest part of the network, but should help bridge emerging partitions. Although more advanced UAV control policies could be developed, the current implementation provides a clear baseline for future research on predictive UAV-assisted fragmentation mitigation.

 The current risk score is not used as the final supervised target. This is an important design decision. If the label were derived directly from the current score, the learning problem would become the replication of a hand-crafted rule. Instead, the target is derived from future graph connectivity. Therefore, the model must learn how current mobility, topology, UAV coverage, and traffic conditions evolve into future fragmentation. This makes the dataset more meaningful for predictive modeling.

 \subsection{Limitations}
 \label{subsec:limitations}

 The current implementation provides a reproducible and extensible foundation, but several limitations should be acknowledged. First, the connectivity graph is derived from an effective range model. This simplifies the wireless channel and does not fully capture fading, interference, obstacles, antenna patterns, or channel load. Second, the mobility model is generated internally using a controlled two-lane highway model rather than being imported from large-scale real-world mobility traces. This improves reproducibility but limits geographical realism. Third, the proactive UAV policy is heuristic and does not optimize energy consumption, collision avoidance, flight constraints, or regulatory restrictions. Fourth, the future labeling strategy is based on graph-level fragmentation thresholds. Although these thresholds are interpretable, they may be adjusted depending on application requirements.

 These limitations define clear directions for extension. Future versions of the generator can integrate SUMO mobility traces, more detailed propagation models, multi-channel communication, UAV energy constraints, reinforcement learning-based UAV positioning, and additional labels such as expected PDR degradation, route failure probability, or safety-message unavailability.

 \subsection{Conclusion}
 \label{subsec:conclusion}

 This paper presented a reproducible UAV-assisted VANET dataset generator for short-term fragmentation risk analysis in Intelligent Transportation Systems. The proposed framework simulates a two-lane highway scenario with vehicles moving in opposite directions, UAVs acting as aerial relays, optional accident zones, and traffic bursts. The generator periodically extracts mobility, topology, UAV coverage, and communication-window features, then assigns each sample a future fragmentation label based on the network state observed after a configurable prediction horizon.

 The main contribution of the work is not a single machine learning model, but a reusable data generation framework that enables supervised learning for proactive VANET fragmentation prediction. By combining scenario diversity, graph-based connectivity analysis, UAV coverage metrics, traffic-window measurements, and future-oriented labeling, the generator provides a practical foundation for evaluating machine learning models and UAV-assisted mitigation strategies. Future work will extend the framework toward realistic SUMO-based mobility, richer wireless channel models, larger multi-seed datasets, and learning-based UAV control policies capable of reducing fragmentation before packet delivery performance collapses.



\section*{Acknowledgements}
The authors would like to acknowledge the Université du Québec en Outaouais (UQO) for its institutional support.

\section*{Funding}
The authors received no specific funding for this study.

\section*{Author Contributions}
\textbf{Bappa Muktar:} Conceptualization, Methodology, Software, Investigation, and Writing---original draft. 
\textbf{Justin Moskolaï Ngossaha: }Methodology and Writing---review \& editing. 
\textbf{Adama Nouboukpo:} Writing---review \& editing.

\section*{Availability of Data and Materials}
The data that support the findings of this study are available from the Corresponding Author, B.M., upon reasonable request.

\section*{Ethics Approval}
Ethics approval was not required for this study because it is based exclusively on synthetic simulation data and does not involve human participants, animals, or personal data.

\section*{Conflicts of Interest}
The authors declare no conflicts of interest to report regarding the present study.

\clearpage

\section*{Abbreviations}
\noindent The following abbreviations are used in this manuscript:\\

\begin{tabular}{ll}
AODV & Ad hoc On-Demand Distance Vector \\
AI & Artificial Intelligence \\
CNN & Convolutional Neural Network \\
CRKN & Canadian Research Knowledge Network \\
DDoS & Distributed Denial of Service \\
FANET & Flying Ad Hoc Network \\
GRU & Gated Recurrent Unit \\
IEEE & Institute of Electrical and Electronics Engineers \\
IP & Internet Protocol \\
IPv4 & Internet Protocol version 4 \\
ITS & Intelligent Transportation Systems \\
LCC & Largest Connected Component \\
NLCC & Normalized Largest Connected Component \\
ns-3 & Network Simulator 3 \\
OMNeT++ & Objective Modular Network Testbed in C++ \\
PDR & Packet Delivery Ratio \\
QoS & Quality of Service \\
RSU & Roadside Unit \\
SDN & Software-Defined Networking \\
SD-VANET & Software-Defined Vehicular Ad Hoc Network \\
SUMO & Simulation of Urban Mobility \\
UAV & Unmanned Aerial Vehicle \\
UDP & User Datagram Protocol \\
V2I & Vehicle-to-Infrastructure \\
V2V & Vehicle-to-Vehicle \\
V2X & Vehicle-to-Everything \\
VANET & Vehicular Ad Hoc Network \\
\end{tabular}

\bibliographystyle{unsrtnat}
 \bibliography{references}  

@article{pawar2024,
  title={Intelligent transportation system with 5G vehicle-to-everything ({V2X}): Architectures, vehicular use cases, emergency vehicles, current challenges, and future directions},
  author={Pawar, Vaishali and Zade, Nilima and Vora, Deepali and Khairnar, Vaishali and Oliveira, Aurenice and Kotecha, Ketan and Kulkarni, Ambarish},
  journal={IEEE Access},
  volume={12},
  pages={183937--183960},
  year={2024},
  publisher={IEEE},
  note={\doi{10.1109/ACCESS.2024.3506815}}
}

@article{shahwani2022,
  title={A comprehensive survey on data dissemination in Vehicular Ad Hoc Networks},
  author={Shahwani, Hamayoun and Shah, Syed Attique and Ashraf, Muhammad and Akram, Muhammad and Jeong, Jaehoon Paul and Shin, Jitae},
  journal={Vehicular Communications},
  volume={34},
  pages={100420},
  year={2022},
  publisher={Elsevier},
  note={\doi{10.1016/j.vehcom.2021.100420}}
}

@article{neelakantan2013,
  title={Connectivity analysis of vehicular ad hoc networks from a physical layer perspective},
  author={Neelakantan, P. C. and Babu, A. V.},
  journal={Wireless Personal Communications},
  volume={71},
  number={1},
  pages={45--70},
  year={2013},
  publisher={Springer},
  note={\doi{10.1007/s11277-012-0795-z}}
}

@article{ashraf2024,
  title={Integrating unmanned aerial vehicles ({UAVs}) with vehicular ad hoc networks ({VANETs}): Architectures, applications, and opportunities},
  author={Ashraf, Muhammad Mansoor and Boudjit, Saadi and Zeadally, Sherali and Bahloul, Nour El Houda and Bashir, Nouman},
  journal={Computer Networks},
  volume={255},
  pages={110873},
  year={2024},
  publisher={Elsevier},
  note={\doi{10.1016/j.comnet.2024.110873}}
}

@article{jain2024,
  title={An efficient multi-objective {UAV}-assisted {RSU} deployment ({MOURD}) scheme for {VANET}},
  author={Jain, Samkit and Jain, Vinod Kumar and Mishra, Subodh},
  journal={Ad Hoc Networks},
  volume={163},
  pages={103598},
  year={2024},
  publisher={Elsevier},
  note={\doi{10.1016/j.adhoc.2024.103598}}
}

@article{yuan2022,
  title={Machine learning for next-generation intelligent transportation systems: A survey},
  author={Yuan, Tingting and da Rocha Neto, Wilson and Rothenberg, Christian Esteve and Obraczka, Katia and Barakat, Chadi and Turletti, Thierry},
  journal={Transactions on Emerging Telecommunications Technologies},
  volume={33},
  number={4},
  pages={e4427},
  year={2022},
  publisher={Wiley Online Library},
  note={\doi{10.1002/ett.4427}}
}

@incollection{riley2010,
  title={The {ns-3} network simulator},
  author={Riley, George F. and Henderson, Thomas R.},
  booktitle={Modeling and Tools for Network Simulation},
  pages={15--34},
  year={2010},
  publisher={Springer},
  note={\doi{10.1007/978-3-642-12331-3_2}}
}

@inproceedings{behrisch2011,
  title={{SUMO}--simulation of urban mobility: An overview},
  author={Behrisch, Michael and Bieker, Laura and Erdmann, Jakob and Krajzewicz, Daniel},
  booktitle={Proceedings of SIMUL 2011, the Third International Conference on Advances in System Simulation},
  year={2011},
  publisher={ThinkMind}
}

@article{dutta2024,
  title={A comprehensive review of recent developments in {VANET} for traffic, safety \& remote monitoring applications},
  author={Dutta, Arijit and Samaniego Campoverde, Luis Miguel and Tropea, Mauro and De Rango, Floriano},
  journal={Journal of Network and Systems Management},
  volume={32},
  number={4},
  pages={73},
  year={2024},
  publisher={Springer},
  note={\doi{10.1007/s10922-024-09853-5}}
}

@article{okello2025,
  title={Connectivity Analysis in {VANETs} with Dynamic Ranges},
  author={Okello, Kenneth and Mwangi, Elijah and El-Malek, Ahmed H Abd},
  journal={Telecom},
  volume={6},
  number={2},
  pages={33},
  year={2025},
  publisher={MDPI},
  note={\doi{10.3390/telecom6020033}}
}

@article{gu2024,
  title={Cluster-based {RSU} deployment strategy for vehicular ad hoc networks with integration of communication, sensing and computing},
  author={Gu, Xinrui and Wang, Shengfeng and Wei, Zhiqing and Feng, Zhiyong},
  journal={Journal of Information and Intelligence},
  volume={2},
  number={4},
  pages={325--338},
  year={2024},
  publisher={Elsevier},
  note={\doi{10.1016/j.jiixd.2024.02.002}}
}

@article{zear2025,
  title={Network partitioning problem and {UAVs}' integration for efficient connectivity restoration: A systematic review},
  author={Zear, Aditi and Ranga, Virender and Bhushan, Kriti},
  journal={International Journal of Communication Systems},
  volume={38},
  number={3},
  pages={e6107},
  year={2025},
  publisher={Wiley Online Library},
  note={\doi{10.1002/dac.6107}}
}

@inproceedings{bouchrit2023,
  title={Flying to the rescue: {UAV}-assisted urgent alert transmission in {VANET}},
  author={Bouchrit, Leila and Zairi, Sajeh and Msadaa, Ikbal C and Dhraief, Amine and Drira, Kahlil},
  booktitle={2023 IEEE International Conference on Enabling Technologies: Infrastructure for Collaborative Enterprises (WETICE)},
  pages={1--6},
  year={2023},
  publisher={IEEE},
  note={\doi{10.1109/WETICE57085.2023.10477830}}

}

@article{sethu2025,
  title={Integrated {VANETs} and {FANETs} Driven Multimodal Smart Transportation System for Delivery},
  author={Sethu Narayanan K and Manu Prakash M and Vetrivelan P and Ajeyprasaath KB},
  journal={IETE Journal of Research},
  volume={71},
  number={2},
  pages = {492--498},
  year={2025},
  publisher = {Taylor \& Francis},
  note={\doi{10.1080/03772063.2024.2420741}}
}

@article{ye2025,
  title={Improving Transmission in Integrated Unmanned Aerial Vehicle--Intelligent Connected Vehicle Networks with Selfish Nodes Using Opportunistic Approaches},
  author={Ye, Meixin and Zhou, Zhenfeng and Zhu, Lijun and Huang, Fanghui and Li, Tao and Wang, Dawei and Jin, Yi and He, Yixin},
  journal={Drones},
  volume={9},
  number={1},
  pages={12},
  year={2025},
  publisher={MDPI},
  note={\doi{10.3390/drones9010012}}
}

@article{rashid2023,
  title={An adaptive real-time malicious node detection framework using machine learning in vehicular ad-hoc networks ({VANETs})},
  author={Rashid, Kanwal and Saeed, Yousaf and Ali, Abid and Jamil, Faisal and Alkanhel, Reem and Muthanna, Ammar},
  journal={Sensors},
  volume={23},
  number={5},
  pages={2594},
  year={2023},
  publisher={MDPI},
  note={\doi{10.3390/s23052594}}
}

@article{bodkhe2025,
  title={Indian SUMO traffic scenario-based misbehaviour detection dataset for connected vehicles},
  author={Bodkhe, Umesh and Tanwar, Sudeep},
  journal={Multimodal Transportation},
  volume={4},
  number={1},
  pages={100148},
  year={2025},
  publisher={Elsevier},
  note={\doi{10.1016/j.multra.2024.100148}}
}

@article{anjali2025,
  title={X-GEVON-a novel explainable intelligent network to detect the multiple attacks in vanet systems},
  author={Anjali, Thuvva and Goyal, Rajeev and Balaji, GN},
  journal={Discover Computing},
  volume={28},
  number={1},
  pages={185},
  year={2025},
  publisher={Springer},
  note={\doi{10.1007/s10791-025-09696-x}}
}

@article{ercan2022,
  title={Misbehavior detection for position falsification attacks in {VANETs} using machine learning},
  author={Ercan, Secil and Ayaida, Marwane and Messai, Nadhir},
  journal={IEEE Access},
  volume={10},
  pages={1893--1904},
  year={2022},
  publisher={IEEE},
  note={\doi{10.1109/ACCESS.2021.3136706}}
}

@article{setia2024,
  title={Securing the road ahead: Machine learning-driven {DDoS} attack detection in {VANET} cloud environments},
  author={Setia, Himanshu and Chhabra, Amit and Singh, Sunil K and Kumar, Sudhakar and Sharma, Sarita and Arya, Varsha and Gupta, Brij B and Wu, Jinsong},
  journal={Cyber Security and Applications},
  volume={2},
  pages={100037},
  year={2024},
  publisher={Elsevier},
  note={\doi{10.1016/j.csa.2024.100037}}
}

@article{polat2024,
  title={Hybrid {AI}-powered real-time distributed denial of service detection and traffic monitoring for software-defined-based vehicular Ad Hoc networks: A new paradigm for securing intelligent transportation networks},
  author={Polat, Onur and Oyucu, Saadin and T{\"u}rko{\u{g}}lu, Muammer and Polat, H{\"u}seyin and Aksoz, Ahmet and Yard{\i}mc{\i}, Fahri},
  journal={Applied Sciences},
  volume={14},
  number={22},
  pages={10501},
  year={2024},
  publisher={MDPI},
  note={\doi{10.3390/app142210501}}
}

@article{haydari2022,
  title={{RSU}-Based Online Intrusion Detection and Mitigation for {VANET}},
  author={Haydari, Ammar and Yilmaz, Yasin},
  journal={Sensors},
  volume={22},
  number={19},
  pages={7612},
  year={2022},
  publisher={MDPI},
  note={\doi{10.3390/s22197612}}
}

@article{gopi2022,
  title={Intelligent {DoS} attack detection with congestion control technique for {VANETs}},
  author={Gopi, R and Mathapati, Mahantesh and Prasad, B and Sultan, Ahmad and Al-Wesabi, Fahd N and Alohali, Manal Abdullah and Hilal, Anwer Mustafa},
  journal={Computers, Materials, \& Continua},
  volume={72},
  number={1},
  pages={141--156},
  year={2022},
  publisher={Tech Science Press},
  note={\doi{10.32604/cmc.2022.023306}}
}

@article{elsadig2025,
  title={Connected vehicles security: A lightweight machine learning model to detect {VANET} attacks},
  author={Elsadig, Muawia A and Altigani, Abdelrahman and Mohamed, Yasir and Mohamed, Abdul Hakim and Kannan, Akbar and Bashir, Mohamed and Adiel, Mousab AE},
  journal={World Electric Vehicle Journal},
  volume={16},
  number={6},
  pages={324},
  year={2025},
  publisher={MDPI},
  note={\doi{10.3390/wevj16060324}}
}






\clearpage

\appendix
\section{Comparative Positioning of Related Work}
\label{app:related_work_positioning}

This appendix provides the detailed comparative table supporting the positioning analysis presented in Section~\ref{sec2.5}. The table summarizes the main focus, methodological scope, UAV support, and relevance of each reviewed study with respect to the proposed UAV-assisted VANET dataset generator.

\setcounter{table}{0}
\renewcommand{\thetable}{A\arabic{table}}

\begin{sidewaystable*}[p]
\centering
\scriptsize
\setlength{\tabcolsep}{3pt}
\renewcommand{\arraystretch}{1.05}

\caption{Comparative positioning of related work and the proposed UAV-assisted VANET dataset generator.}
\label{tab:appendix_related_work_positioning}

\begin{tabularx}{\textheight}{
@{}
P{0.055\textheight}
>{\hsize=0.75\hsize\linewidth=\hsize}Y
>{\hsize=0.95\hsize\linewidth=\hsize}Y
C{0.065\textheight}
>{\hsize=1.30\hsize\linewidth=\hsize}Y
@{}
}
\toprule
\textbf{Ref.} &
\textbf{Main focus} &
\textbf{Methodological scope / data source} &
\textbf{UAV support} &
\textbf{Positioning with respect to the proposed work} \\
\midrule

\cite{pawar2024} &
5G V2X architectures, emergency vehicle communication, cooperative driving, and ITS use cases. &
Review of 5G-enabled V2X architectures and vehicular communication requirements. &
No &
Supports the need for reliable and low-latency vehicular communication, but does not address fragmentation prediction or supervised dataset generation. \\

\cite{ashraf2024} &
UAV integration with VANETs, including architectures, applications, and open challenges. &
Review and taxonomy of UAV--VANET integration approaches. &
Yes &
Establishes UAVs as promising aerial relays for VANETs, but does not propose a reproducible generator for fragmentation risk forecasting. \\

\cite{dutta2024} &
VANET connectivity, safety applications, traffic monitoring, and remote sensing services. &
Review of VANET communication technologies, routing, clustering, monitoring, and service support mechanisms. &
No &
Highlights the importance of connectivity-aware VANET design, but does not provide a reusable simulation-based dataset generator for future fragmentation risk prediction. \\

\cite{okello2025} &
VANET connectivity analysis under dynamic communication ranges. &
Analytical connectivity modeling considering variable radio range, fading, and communication uncertainty. &
No &
Shows that VANET fragmentation cannot be analyzed only through graph proximity; however, it does not generate future-labeled fragmentation datasets. \\

\cite{gu2024} &
Cluster-based RSU deployment integrating communication, sensing, and computing. &
Road-network-based RSU deployment model using hierarchical placement criteria. &
No &
Demonstrates the impact of road topology and infrastructure placement on connectivity, but focuses on fixed RSUs rather than UAV-assisted predictive dataset generation. \\

\cite{zear2025} &
Network partitioning and UAV-based connectivity restoration. &
Systematic review of partition detection, recovery, and UAV-assisted reconnection mechanisms. &
Yes &
Treats fragmentation as a network partitioning problem, but focuses on recovery after disconnection rather than predicting future fragmentation before performance collapse. \\

\cite{bouchrit2023} &
UAV-assisted urgent alert transmission in sparse rural highway VANETs. &
UAV--VANET highway communication scenario for urgent message dissemination. &
Yes &
Demonstrates that UAV relays can mitigate sparse highway connectivity gaps, but does not construct labeled datasets for short-term fragmentation prediction. \\

\cite{sethu2025} &
Integrated VANET--FANET multimodal smart transportation for delivery services. &
System-level integration of vehicular and flying ad hoc networks for smart transportation services. &
Yes &
Reflects the convergence of vehicular and aerial networks, but remains service-oriented rather than dataset-oriented for fragmentation risk analysis. \\

\cite{ye2025} &
Transmission improvement in integrated UAV--Intelligent Connected Vehicle networks. &
UAV-assisted vehicular delay-tolerant network model with opportunistic transmission. &
Yes &
Addresses intermittent connectivity and delivery performance, but does not provide reusable supervised learning data for future fragmentation prediction. \\

\cite{rashid2023} &
Adaptive real-time malicious node detection in VANETs. &
OMNeT++ and SUMO-based VANET simulation traces for machine learning-based malicious node detection. &
No &
Shows the usefulness of simulator-derived VANET datasets for supervised learning, but targets security detection rather than topology degradation or fragmentation forecasting. \\

\cite{bodkhe2025} &
SUMO-based misbehavior detection dataset for connected vehicles in an Indian ITS scenario. &
Ahmedabad SUMO Traffic scenario and AhmST dataset for false data injection and misbehavior classification. &
No &
Improves geographical diversity and reproducibility in VANET datasets, but focuses on misbehavior detection rather than UAV-assisted future fragmentation risk. \\

\cite{anjali2025} &
Explainable intelligent network for detecting multiple VANET attacks. &
SUMO, OMNeT++, and Python-based dataset construction with deep learning and explainability. &
No &
Combines dataset generation, learning, and interpretability, but remains attack-centered and does not define future fragmentation labels. \\

\cite{ercan2022} &
Position falsification attack detection in VANETs. &
VANET message and mobility-related features for machine learning-based position falsification detection. &
No &
Highlights the value of engineered mobility and position features, but does not address connectivity fragmentation or future-oriented risk labeling. \\

\cite{setia2024} &
DDoS attack detection in VANET cloud environments. &
VANET cloud network-flow data analyzed using machine learning and fuzzification. &
No &
Demonstrates the effectiveness of learning-based traffic analysis, but does not consider UAV coverage, graph fragmentation indicators, or future disconnection labels. \\

\cite{polat2024} &
Hybrid AI-powered DDoS detection and traffic monitoring in SD-VANETs. &
Software-defined VANET architecture with real-time DDoS detection using 1D-CNN and decision tree models. &
No &
Shows the value of real-time AI-based VANET monitoring, but focuses on cyberattack detection rather than future fragmentation prediction. \\

\cite{haydari2022} &
RSU-based online intrusion detection and mitigation in VANETs. &
Traffic simulation and infrastructure-based monitoring through RSU-supported anomaly detection. &
No &
Uses infrastructure monitoring to improve VANET security, whereas the proposed work emphasizes UAV-assisted topology monitoring and future fragmentation labeling. \\

\cite{gopi2022} &
DoS attack detection with congestion control in VANETs. &
VANET simulation environment with optimization and GRU-based DoS detection. &
No &
Links attacks and congestion to communication degradation, but does not provide a modular future-labeled dataset generator for fragmentation risk. \\

\cite{elsadig2025} &
Lightweight machine learning model for VANET attack detection. &
Simulated VANET security dataset with feature selection, class balancing, and lightweight classification. &
No &
Reinforces the importance of preprocessing and feature reduction, but remains security-oriented rather than topology-aware fragmentation forecasting. \\

\textbf{Proposed work} &
\textbf{UAV-assisted VANET dataset generation for short-term fragmentation risk analysis.} &
\textbf{ns-3-based UAV-assisted VANET simulation with configurable highway profiles, UAV policies, mobility disturbances, and traffic windows.} &
\textbf{Yes} &
\textbf{Provides a reusable dataset generator that jointly extracts mobility, topology, UAV coverage, and communication-window features, then assigns future fragmentation labels using a configurable prediction horizon.} \\

\bottomrule
\end{tabularx}

\vspace{3pt}
\begin{minipage}{0.96\textheight}
\scriptsize
\textit{Note:} RSU: Roadside Unit; UAV: Unmanned Aerial Vehicle; VANET: Vehicular Ad Hoc Network; FANET: Flying Ad Hoc Network; ITS: Intelligent Transportation Systems; SD-VANET: Software-Defined VANET.
\end{minipage}

\end{sidewaystable*}

\clearpage

\section{Object-Oriented Architecture of the Proposed Generator}
\label{app:oop_architecture}

This appendix provides the detailed Doxygen-generated object-oriented architecture diagram of the proposed UAV-assisted VANET dataset generator. The diagram complements Section~\ref{subsec:oo_architecture} by showing the relationships among the main implementation components.

\begin{figure*}[!p]
\centering
\includegraphics[width=1.1\textwidth]{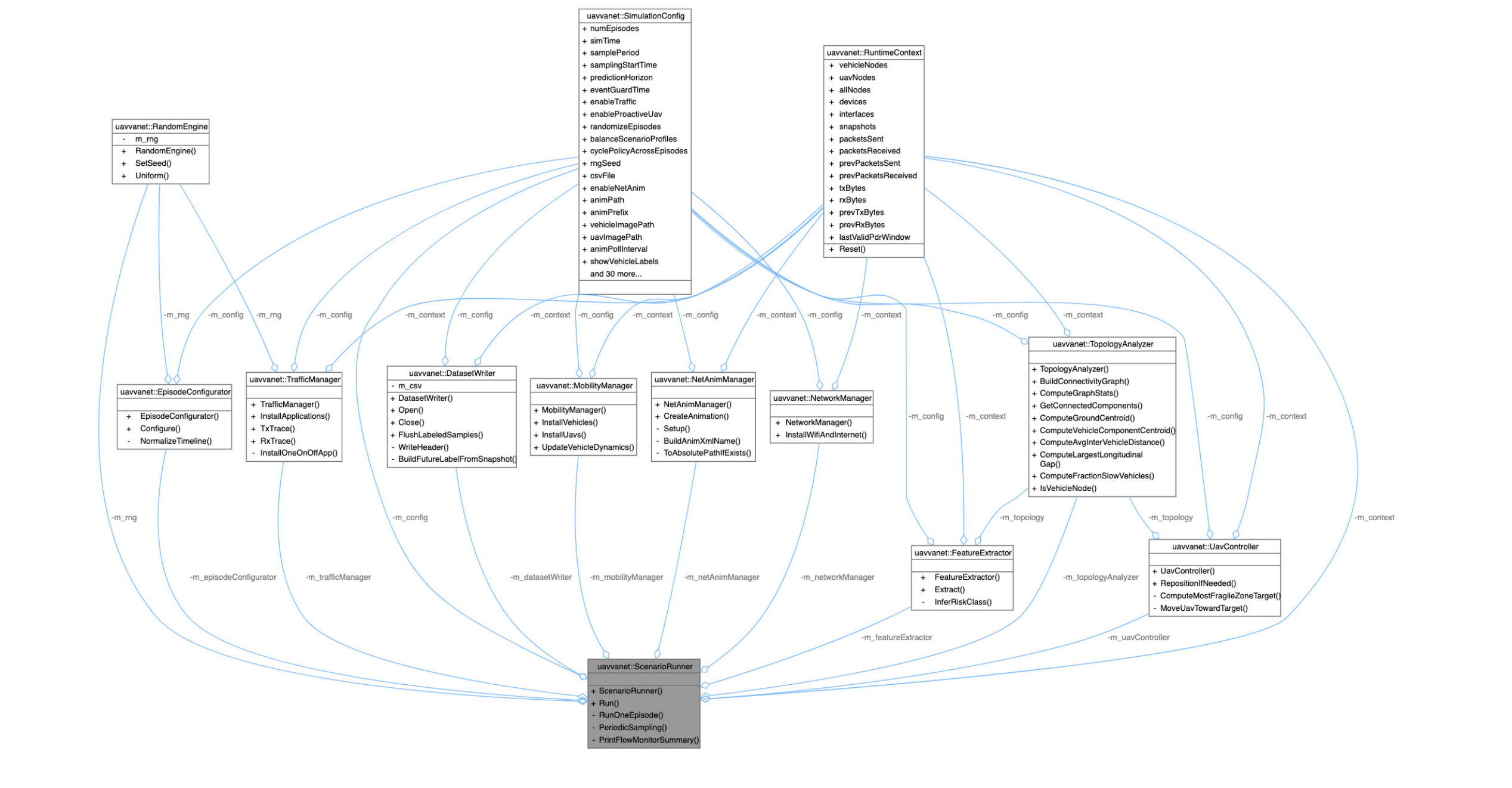}
\caption{Object-oriented structure of the proposed UAV-assisted VANET dataset generator. The diagram highlights the central role of \texttt{ScenarioRunner} in coordinating episode configuration, mobility installation, network setup, traffic generation, feature extraction, UAV control, and dataset writing.}
\label{fig:oop_architecture}
\end{figure*}

\end{document}